%% file: Paper2.tex
\documentclass[useAMS,usenatbib]{mn2e}

\usepackage{amsmath}
\usepackage{relsize}
\usepackage{epstopdf}
\usepackage{graphicx}
\usepackage{graphics}
\usepackage{multirow}
\usepackage{booktabs}
\usepackage{supertabular}
\usepackage{lscape}
\usepackage{longtable}
\usepackage[figuresright]{rotating}
\usepackage{multicol}
\usepackage{natbib}
\usepackage{rotating}
\usepackage{epsfig}
\usepackage{amsfonts}
\usepackage{threeparttable}
\usepackage[english]{babel}
\usepackage{textcomp}
\usepackage{natbib}
\usepackage{subfigure}
\usepackage{tabularx}
\usepackage{pifont}
\usepackage{float}
\newcommand{\tm}{\ding{52}}
\newcommand{\xm}{\ding{54}}

\usepackage{amssymb} 
\usepackage{url} 

\newcolumntype{R}[1]{>{\raggedleft\arraybackslash}p{#1}}
\newcolumntype{C}[1]{>{\centering\arraybackslash}p{#1}}
\newcolumntype{L}[1]{>{\raggedright\arraybackslash}p{#1}}

\title[Searching for New Nearby Young Moving Groups in the Northern Hemisphere]{A Kinematically Unbiased Search for Nearby Young Stars in the Northern Hemisphere Selected Using SuperWASP Rotation Periods}
\author[A. S. Binks et al.]{A. S. Binks\thanks{E-mail: a.s.binks@keele.ac.uk}, R. D. Jeffries and P. F. L. Maxted\\
Astrophysics Group, Keele University, Keele, Staffordshire ST5 5BG}
\begin{document}

\date{Accepted. Received; in original form}

\pagerange{\pageref{firstpage}--\pageref{lastpage}} \pubyear{2015}

\maketitle

\label{firstpage}

\begin{abstract}
We present a kinematically-unbiased search to identify young, nearby low-mass members of kinematic moving groups (MGs). Objects with both rotation periods shorter than 5 days in the SuperWASP All-Sky Survey and X-ray counterparts in the ROSAT All-Sky Survey were chosen to create a catalog of several thousand rapidly-rotating, X-ray active FGK stars. These objects are expected to be either young single stars or tidally-locked spectroscopic binaries. We obtained optical spectra for a sub-sample of 146 stars to determine their ages and kinematics, and in some cases repeat radial velocity (RV) measurements were used to identify binarity. Twenty-six stars are found to have lithium abundances consistent with an age of $\leq 200$\,Myr, and show no evidence for binarity and in most cases measurements of H$\alpha$ and $v\sin i$ support their youthful status. Based on their youth, their radial velocities and estimates of their 3-dimensional kinematics, we find 11 objects that may be members of known MGs, 8 that do not appear associated with any young MG and a further 7 that are close to the kinematics of the recently proposed ``Octans-Near'' MG, and which may be the first members of this MG found in the northern hemisphere. The initial search mechanism was $\sim 18$ per cent efficient at identifying likely-single stars younger than 200\,Myr, of which 80 per cent were early-K spectral types.
\end{abstract}


\section{Introduction}\label{S_Intro}

In a series of publications in the 1960s, Olin Eggen hypothesised the existence of a ``Local Association'' of co-moving, early-type stars in the solar vicinity \citep{1961a_Eggen, 1965a_Eggen}. This association (also known as the Pleiades moving group) included members of the Pleiades, $\alpha$~Persei and IC~2602 open clusters, and roughly one-third of the B7-A0 stars within 300pc \citep{1983a_Eggen}.

Following Eggen's hypothesis, detailed observations of several young, chromospherically active, fast-rotating, late-type stars led to the suggestion that they too were part of the the Local Association (e.g. AB Dor, PZ Tel, \citealt{1988a_Innis}; BO Mic, \citealt{1993a_Anders}; LO Peg, \citealt{1994a_Jeffries}). Follow-up spectroscopy of optical counterparts to coronally active late-type stars found in EUV and X-ray surveys found that a large fraction were co-moving with the Local Association \citep{1993a_Jeffries}; many were as young or younger than the Pleiades ($\sim 100$\,Myr) as evidenced by the presence of lithium in their photospheres, which is otherwise burned rapidly during the pre main sequence in low mass stars \citep{1995a_Jeffries}. Subsequent work (e.g. \citealt{2000a_Zuckerman, 2000a_Torres, 2001a_Zuckerman, 2001a_Montes, 2004a_Zuckerman}) suggested that the Local Association has kinematic and spatial substructure, consisting of several different co-eval, co-moving streams of stars at a range of young ages, that have been collectively termed ``nearby young moving groups'' (hereafter referred to as MGs, \citealt{2004a_Zuckerman, 2008a_Torres}). The origins of these MGs is still an open question. They may be the result of the dissolution of young open clusters, with members of each MG sharing a common birthplace as well as common kinematics. For instance, \cite{2013a_De_Silva} found that stars in the Argus MG and the IC~2391 open cluster shared common chemical abundances and kinematics. However, an abundance analysis of a population of objects in the AB Doradus MG (ABDMG) by \cite{2013a_Barenfeld} indicates that approximately half of the previously suggested members do \textit{not} share a similar chemical composition, arguing against a common origin.

Finding young stars in the solar neighbourhood is important because they represent some of the best observational targets for understanding the early evolution of stars and their surrounding circumstellar environments and planetary systems (e.g. \citealt{2013a_Dent, 2014a_Brandt, 2015a_Bowler}).  They are much closer than their equivalents in young clusters and star forming regions, offering advantages both in terms of sensitivity and spatial resolution. If stars can be linked to particular coeval MGs, then their ages can reasonably be assumed similar to that MG as a whole. At ages of 10--100\,Myr gas giant planets around MG members are expected to be much more luminous than in older systems, and young stars are frequently surrounded by debris discs that may evidence the formation of terrestrial planets or provide diagnostic indicators of unseen planets. Lower mass MG members potentially provide even better targets to investigate planets and circumstellar environments because their lower luminosities enhance the contrast with giant planets of a given mass. Examples of work that exploits the youth and proximity of MG members includes the high contrast infrared imaging detection of multiple planets surrounding the 30 Myr old A-type dwarf, HR~8799, a member of the Columba MG \citep{2008a_Marois, 2010a_Marois}, and the identification of a planet around the A0 star $\beta$~Pic; the eponymous member of the $\beta$~Pictoris~MG (BPMG, \citealt{2010a_Lagrange}).

Much work has focused on finding new low-mass MG members using both kinematic selection and kinematically unbiased surveys. Kinematic selection may be efficient at discovering new members of known MGs, but precludes the discovery of nearby, young objects that are not members of these groups. Examples include the proper-motion selected searches reported by Schlieder, Lepine \& Simon (2010, 2012) or the work of Malo et al. (2013, 2014) and Gagn\'{e} et al. (2014, 2015) who used both positions and Galactic velocities to assign probabilities of membership to new candidate members of several known MGs. 

\nocite{2010a_Schlieder}
\nocite{2012a_Schlieder}
\nocite{2013a_Malo}
\nocite{2014a_Malo}
\nocite{2014a_Gagne}
\nocite{2015a_Gagne}

Kinematically unbiased searches are possible but less efficient. For example, from an initial sample of 405 late-type stars within 25\,pc of the Sun, \cite{2010a_Maldonado} found only 6 per cent that may be candidate members of known MGs based on their space motions. A more focused approach is to pre-select stars which are likely to be young based on their magnetic activity. Young stars are magnetically active as a result of their fast rotation, convective envelopes and consequent dynamo-generated magnetic fields; this activity is manifested as chromospheric and coronal emission that can be detected via optical emission lines or UV and X-ray flux. Examples of this approach can be found in the earlier works of \cite{1995a_Jeffries} and \cite{2001a_Montes}, but more recently \cite{2006a_Torres}, \cite{2009a_Lepine}, \cite{2009a_da_Silva} and \cite{2011a_Shkolnik}. Notably, \cite{2012a_Shkolnik} pre-selected a sample of nearby, X-ray active M-dwarfs, finding many new MG members but also finding that about 50 per cent of the young M-dwarfs could not be assigned to any of the currently known MGs.

Here we describe a new, kinematically unbiased method to select young stars that relies on the fact that stellar rotation is strongly age dependent. At young ages, a large fraction of low-mass stars have fast rotation rates (rotation periods less than a few days -- e.g. \citealt{1996a_Patten, 1998a_Krishnamurthi}). Angular momentum loss due to magnetised stellar winds leads to spin down on a mass-dependent timescale, ranging from $\simeq 50$\,Myr for G-stars to hundreds of Myr for M-dwarfs (see for example \citealt{2003a_Barnes}). \cite{2010a_Messina} report rotation periods of the order of several days or less for many MG members, confirming that a selection based on a short rotation period is likely to favour young stars, although may be contaminated by members of older, tidally-locked short period binary systems. Confirming the youth of candidates requires high resolution spectroscopy.  Young F- ,G- and K-stars should have large abundances of lithium in their photospheres (e.g. \citealt{2005a_Sestito, 2006a_Jeffries}) and multiple radial velocity measurements can be used to identify short-period binary systems.

In this paper we report our initial efforts to find young, nearby stars using a parent sample of fast-rotating, active stars selected from the union of the SuperWASP transiting planet survey and the ROSAT X-ray all-sky survey. These candidates were followed up with high resolution spectroscopy at the Nordic Optical Telescope (NOT) and Isaac Newton Telescope (INT). In $\S$\ref{S_Period} we describe how rotation periods were estimated from the SuperWASP photometry database. The initial candidate selection is described in $\S$\ref{S_Obs} along with details of the spectroscopic observations. In $\S$\ref{S_Data_Analysis} we describe the techniques used to measure radial and rotational velocities, temperatures, chromospheric activity and Li abundances. The multiple methods that were used to constrain the ages of the observed targets are discussed in $\S$\ref{S_Age}, and in $\S$\ref{S_Kin} the space motions of the young, Li-rich targets are calculated and compared to known MGs. We provide a discussion of individual objects in $\S$\ref{S_Indiv} and in $\S$\ref{S_Conclusions} we discuss the potential of a repeat survey, focusing on the efficiency and relative success of this work at identifying kinematic sub-structure in the young sample.

\section{Period determination}\label{S_Period}

The SuperWASP project \citep{2006a_Pollacco} is a wide-field photometric survey for transiting exoplanets that has been operating since 2004. The wide field of view of its two instruments located in La Palma (Spain) and South Africa (one set of eight Canon 200\,mm f-1.8 lenses on each site, covering 482 square degrees for each observatory and backed by high quality $2048 \times 2048$ i-Kon CCD detectors) and typical observing cadence of $\sim$ 10 minutes also make it proficient at identifying many types of stellar variability with timescales from one hour to several weeks. Data are available for tens of millions of objects with brightness in the approximate range $8 < V < 15$ covering most of the sky. An initial catalog of objects was generated by cross-correlating the ROSAT sky-survey (1RXS) and pointed phase (2RXP) catalogues (\citealt{1999a_Voges}, see $\S$\ref{S_Xray} for an analysis of the ROSAT data acquired in this work) with objects in the SuperWASP archive. We created a sample of 5477 stars using the criteria that SuperWASP targets must be within either the $3\sigma$ position uncertainty or 10'' of the X-ray source (whichever was larger), have declinations $>-20^{\circ}$ and contain more than 1000 photometric data points in the archive up to the date 20th July 2010 when the database query was performed.

Periodic variable stars were identified using the Lomb-Scargle periodogram technique described in \cite{2011a_Maxted}. The Lomb-Scargle technique (\citealt{1976a_Lomb, 1982a_Scargle, 1986a_Horne, 2009a_Zechmeister}) searches for significant periodicities in unevenly sampled data. The SuperWASP light-curves are measured over several seasons of observation, each of which typically have $\sim 8000$ unevenly sampled data points. To measure the period, we calculate the normalised power $P_{n}(\omega$) at a given angular frequency, $\omega = 2\pi\nu$. The highest peaks in the calculated power spectrum correspond to candidate periodicities in the time series data. To obtain a solution for the light-curve, a least-squares fit of the sinusoidal function $y_{i} = a\sin({\omega}t_{i}) + b\cos({\omega}t_{i})$ to magnitudes ${i} = 1, 2, . . . ,N$ is found. A power spectrum is obtained based on a chi-squared fit of the light-curve:

\begin{equation}
 P_{n}(\omega) = \frac{\chi_{0}^{2}-\chi^{2}(\omega)}{\chi_{0}^{2}},
\label{E_LC_Fit1}
\end{equation}

where

\begin{equation}
  \chi_{0}^{2} = \sum\frac{m_{i}^{2}}{\sigma_{i}}, \nonumber
\label{E_LC_Fit2}
\end{equation}

and

\begin{equation}
  \chi^{2}(\omega)  = \sum\frac{(m_{i}-y_{i})^{2}}{\sigma_{i}^{2}} \nonumber
\label{E_LC_Fit3}
\end{equation}
\\

To ensure the signal is not a noise artefact, the false alarm probabilities (FAPs) were calculated by making 100 permutations of each light-curve using a bootstrap Monte Carlo technique developed by \cite{2009a_Collier_Cameron}. The FAP related to a given power $P_{n}$ is taken as the fraction of randomised light-curves that have a highest power peak that exceeds $P_{n}$, which is the probability that a peak of a given height is merely caused by statistical variations, i.e. white noise. The spectrum of FAPs was used to estimate the power value in the periodogram for which this probability is 0.1, 1 and 10 per cent. Uncertainties are split into 2 separate error bars: first is the (averaged) measurement error using $\Delta P = \frac{\delta \mu P^{2}}{2}$ (equation 2 in \citealt{2010a_Messina}), where $\delta \mu$ is the finite frequency resolution of the power spectrum and is equal to the full-width half maximum of the highest power peak in the frequency spectrum. The second error bar (where appropriate) is the standard error in 2 or more period measurements. Effects such as differential rotation between seasons may exceed any error bar generated from the width of the peak frequency, therefore errors based on only 1 season are likely to be underestimates (e.g. \citealt{2013a_Reinhold, 2014a_Epstein}).  The error bars of objects with 2 seasons of data also risk under-estimation because of the broad probability distribution of the standard error in 2 measurements.

\begin{figure*}
\begin{center}
\includegraphics[width=0.8\textwidth]{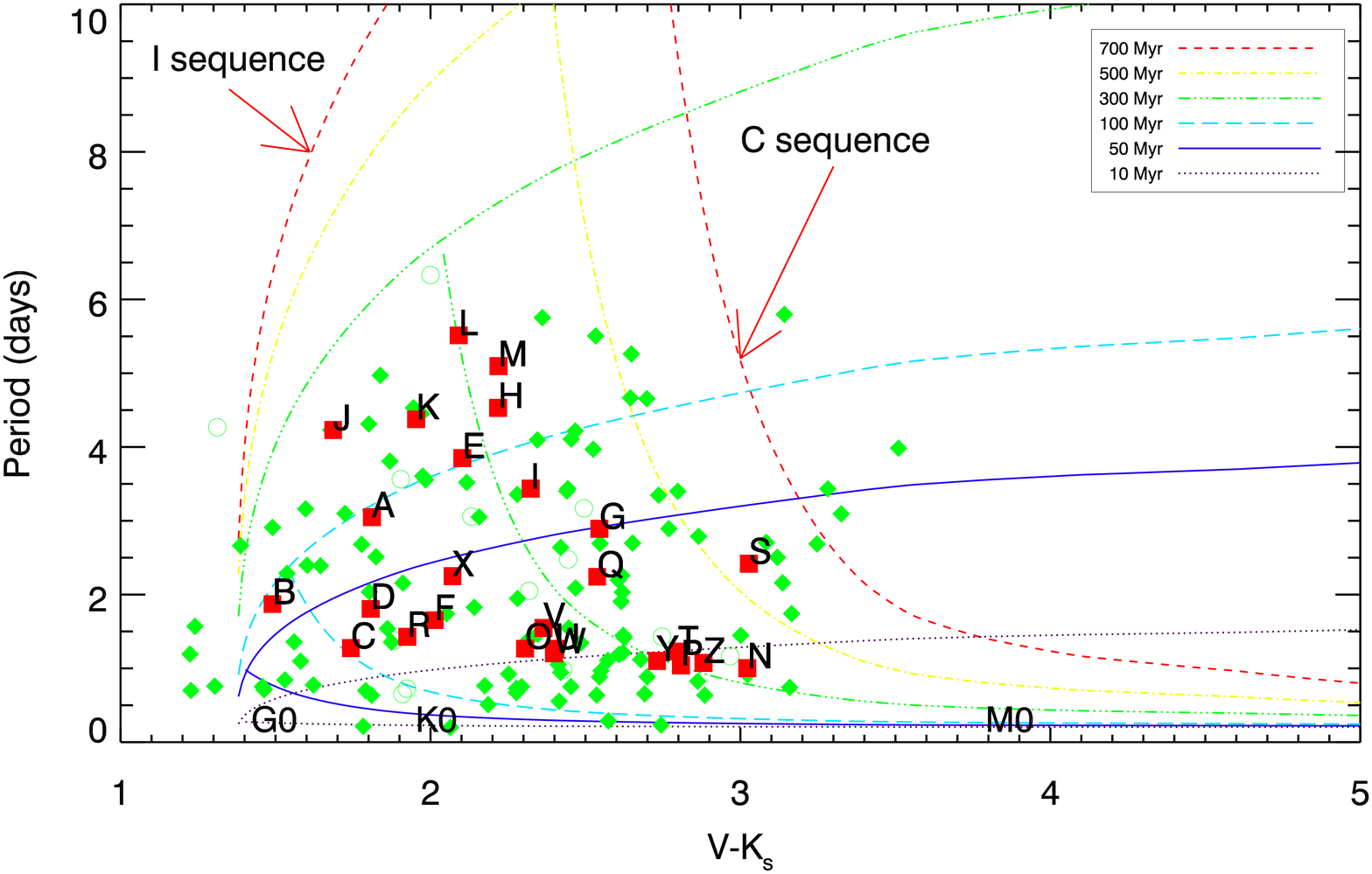}
\end{center}
\begin{flushleft}
 \caption{Rotation periods and $V-K_{\rm s}$ colour for the entire observed sample. Red squares represent 26 objects that were later identified to be likely-young and were considered likely-single stars (referred to as the `likely-young sample', see $\S$\ref{S_Age_Final}). All these objects are labelled from A to Z and their corresponding ROSAT 1RXS names are provided in Table~\ref{T_PX_LYS}. Any objects that had two or more separate RV measurements that varied by more than $5\,{\rm km\,s}^{-1}$ were assumed to be close-in short-period binaries and are denoted here by open green circles. All other objects are designated filled green diamonds. This symbol scheme for all other plots will remain the same throughout this work. The various lines represent the I- and C-sequence gyrochrones proposed by \protect\cite{2003a_Barnes} and calibrated by \protect\cite{2008a_Mamajek}. The method of inferring an age range from the gyrochrones is described in $\S$\ref{S_Gyro} -- note the three objects A, B and L which are used as examples to describe how to fit the age ranges.}
 \label{F_Gyrochrones}
\end{flushleft}
\end{figure*}

Based on all the available periodograms, the periods were given a quality classification. If 3 or more seasons of data for an object resulted in an average period with a standard error smaller than 5 per cent they were graded as `A'. To assess the period determination reliability, we measured the difference in $\chi^{2}$ ($\Delta\chi^{2}$) between the second highest peak in each periodogram to the highest peak. If $\Delta\chi^{2}$, averaged over all observing seasons, was large enough (each light-curve contains $\sim$8000 datapoints, so an average $\Delta\chi^{2} > 1000$ was chosen as our minimum cut-off) and the value of the highest peak was larger than the FAP value at 0.1 per cent, we considered this a reliable period measurement. Objects with 1 or 2 seasons of data and satisfying the average $\Delta\chi^{2}$ criteria, and had a standard error smaller than 5 per cent were graded as `B', whereas objects failing either the $\Delta\chi^{2}$ condition or resulting in large standard errors for their periods were graded `C'.

From the initial catalog of 5477 targets a sub-sample of 146 were chosen for spectroscopic investigation based on inspection of their light-curves, period analysis, and their visibility during the observing run. Rapidly rotating, late-type stars with periods less than 5 days were considered for target selection. A sub-sample of 26 observed targets are listed in Table~\ref{T_PX_LYS} (all other objects are listed in an online supplement). Note that for the purposes of this paper, the light-curves were reanalysed using a larger dataset (see $\S$\ref{S_Period}) and nine of the redetermined periods were subsequently found to have periods $> 5$ days. 

We provide the raw light-curves, the periodograms and corresponding phase-folded light-curves for all objects that we obtained spectroscopy for (see $\S$\ref{S_Sample}) as an online supplement to this paper. A summary of the targets investigated in this paper and their measured periods are provided in Table~\ref{T_PX_LYS} along with their uncertainties and quality classification. Note that the light-curves and the periods quoted in these tables are based on a reanalysis of all the SuperWASP data available up until December 2012, performed after our spectroscopic follow-up. In most cases these agree with the original periods which were based on data up to the 20$^{\rm th}$ July 2010, however there are a handful of significant discrepancies.

The main source of $B$ and $V$ magnitudes are from the AAVSO All-Sky Photometric Survey (APASS, \citealt{2012a_Henden}), which covers the magnitude range $10 < V < 17$. All objects with $V$ magnitudes $< 10$ that were unavailable in the APASS catalog were instead sourced from the NOMAD catalog \citep{2005a_Zacharias}. All $K$ magnitudes are from 2MASS \citep{2003a_Cutri}. Proper-motions are extracted from the PPMXL catalog of positions and proper motions \citep{2010a_Roeser}. $BVK$ photometry and proper-motions are presented in Tables~\ref{T_Temp_EW_Abun_LYS} and~\ref{T_Kin_LYS}, respectively. Figure~\ref{F_Gyrochrones} displays the entire observed sample plotted in terms of their rotation periods and $V-K_{\rm s}$ colour.

\input{P_Xray}

\section{Sample selection and spectroscopic observations}\label{S_Obs}

\subsection{Sample selection}\label{S_Sample}
Prior to selecting targets for spectroscopy, the light-curves for the sample available in the SuperWASP catalog were checked to identify objects most likely to be showing variability in their light-curves as a result of starspot modulation. These were more likely to be objects with several seasonal light-curves. Whilst some objects had consistent periods measured from many seasons of data, with clear sinusoidal behaviour, some of the sample had sparse, poor quality data and sometimes inconsistent period determinations. Because of the range of quality in data the selection process was entirely subjective. Any objects indicating a sharp dip in their light-curve over any part of their phase were flagged as eclipsing binaries and discarded. The light-curve variability amplitudes were restricted to $< 0.2\,$mag and objects with measured periods to within a thousandth of either a half or integer number of days were discarded because they risked incorrect period determinations due to aliasing effects. Priority was given to objects where the profile of the multi-seasonal light-curves significantly changed in shape; a strong indication of starspot migration over the course of several months.

The range of target spectral-types is restricted to the range mid-F to early M (approximately $1.0 < V-K_{\rm s} < 3.5$) for two reasons. First, because lithium is used as a primary youth indicator (see section~\ref{S_Li_EW}), this range contains stars where significant and measurable Li depletion is expected on timescales of $10-100$\,Myr (see \citealt{2010a_Soderblom} and references therein). The hot boundary is where Li-depletion timescales become very long, and poorly understood diffusive or other non-standard mixing processes may become dominant \citep{1986a_Boesgaard}. The cool boundary marks the point where Li depletion occurs extremely rapidly and even very young stars may have depleted their Li (e.g. \citealt{2006a_Jeffries}). Secondly, low-mass MG stars will usually rotate very rapidly; FGK stars typically have spin-down timescales of $\sim 50-100$\,Myr, and M-dwarfs $\sim 300$\,Myr or longer (this is illustrated by the gyrochrones in Figure~\ref{F_Gyrochrones}). Thus fast rotation in M-dwarfs is not necessarily a sign of youth and the targetting of such objects could lead to high contamination with older stars. The basis for choosing our objects in terms of colour and rotation period is discussed in more detail in $\S$\ref{S_Gyro} where we use `gyrochronology' to provide crude age estimates. 

Objects in the SuperWASP catalog may have correlated with ROSAT sources by random chance. To test this, the original positions of all objects observed in the ROSAT catalog were offset by 2' and checked again for neighbouring SuperWASP sources. The ROSAT error circles are about $10-20$'', therefore a search radius of 15'' was used to identify any neighbouring objects in the SuperWASP catalog. Only one object (target W) out of 146 had a neighbouring SuperWASP source subsequent to being shifted by 2'. The chance of random correlation is 0.68 per cent.

An important point is that the nature of the target selection, from an all-sky survey, allows for a \textit{kinematically unbiased} sample, independent of proper-motion or radial velocity (RV) criteria. This provides an opportunity to identify young stars that do not share the galactic space motions of previously identified MGs. The sample is not spatially unbiased, because the SuperWASP survey avoids galactic latitudes between $\pm 20^{\circ}$.

\subsection{Observation strategy}\label{S_Strategy}

We obtained high-resolution multi-echelle spectroscopy of 146 objects over 2 telescope runs using the Fibre-fed Echelle Spectrograph (FIES) on the 2.56\,m Nordic Optical Telescope (NOT) at the Roque de los Muchachos Observatory. Our first observing run of 68 targets lasted 4 consecutive nights from 21 June 2011, and a further 78 targets were observed on 27, 28 and 29 December 2012. FIES was used in medium resolution mode (${\lambda}$/${\Delta\lambda}$ = 46,000). Each target observation was bracketed with short ThAr arc lamp exposures to provide a wavelength calibration. In what follows, objects observed in June 2011 or December 2012 are denoted J11 and D12, respectively (these are listed in the `Date' column in Table~\ref{T_PX_LYS}).

Our observing strategy was to obtain a single observation of a target, reduce the data in real time at the telescope and inspect the spectrum for the presence of the Li\,{\sc i}~6708\AA\ doublet. If Li was clearly present then, if necessary, a further exposure was performed to obtain a signal to noise ratio (SNR) per pixel of $\geq 50$ around the Li line. Objects where Li was detected were observed again on a subsequent night in order to check for short-term RV changes that might betray their close binary nature (see $\S$\ref{S_Bin_Con} for further details on binary contaminants in the survey).

The real time data reduction at the telescope and subsequent reduction at Keele University were accomplished using the \,{\sc FIEStool} package created by Eric Stempels. This package performed bias, blaze and flat field corrections, optimal extraction of the spectra and wavelength calibration using the adjacent arc lamp exposures. With fibre bundle $\#$3, FIES covers a spectral range of $\lambda\lambda$ 3630--7260\AA\ over 78 spectral orders. Observations were also made of several RV and low-activity template stars (selected for their extremely low levels of chromospheric Ca\,{\sc ii} H~and~K emission) over a similar spectral-type range as the targets. These were used to calibrate RVs and projected rotational velocities ($v\sin i$) and are listed in Table~\ref{T_RV_Templates}.

To further identify whether or not objects were short-period binaries, several of the most probable young objects were later re-examined using long-slit spectroscopy on the Isaac Newton Telescope (INT) on the 23$^{\rm rd}$ and 24$^{\rm th}$ March 2013, in combination with the H1800V grating and IDS-235 wide-field camera. This allowed for an RV precision of $\sim 1-2\,{\rm km\,s}^{-1}$ and these measurements were compared to the RVs measured at the NOT. In this work, the RV measurements at the INT are not incorporated into the final averaged RV, rather they are used to check for RV consistency and Li content. All individually measured RVs are presented in column 3 of Table~\ref{T_RVs_LYS} and INT observations are subscripted with an `i'.

\input{RV_Templates}

\section{Data analysis and results}\label{S_Data_Analysis}

\subsection{Equivalent widths (EWs)}\label{S_EW}

Li EWs were estimated from the NOT spectra using the {\sc splot} procedure in {\sc iraf}.\footnote{IRAF is distributed by the National Optical Astronomy Observatory, which is operated by the Association of Universities for Research in Astronomy (AURA) under cooperative agreement with the National Science Foundation.}  Lines were measured by manual integration, where the linear continuum is subtracted and pixels were summed with partial pixels at the ends to obtain a flux. Although the Li line at 6707.8\AA\ actually consists of 2 neighbouring lines at $\lambda\lambda$ 6707.761, 6707.912\AA\ our spectroscopy cannot resolve these lines. There is also a contaminating weak Fe\,{\sc i} spectral line at 6707.441\AA\ blended into the Li resonance line. Where necessary, we use a correction for this blend, $W_{\lambda}{\rm (blend)} = 20(B-V) - {\rm 3}$m\AA\ provided in \cite{1993c_Soderblom}. EW measurement errors were estimated using the procedure highlighted in \cite{1988a_Cayrel_de_Strobel}:

\begin{equation}
 \delta{\rm EW} = 1.6 \times \sqrt{{\rm FWHM} \times p}/{\rm SNR},
\label{E_Cayrel}
\end{equation}
\\
where FWHM refers to the observed full width half maximum (in \AA) of the line, and $p$ is the pixel size (0.113\AA). The SNR was calculated empirically from the rms of fits to continuum regions.

Measurements of the 6563\AA\ H$\alpha$ line EW were made by direct integration above a surrrounding pseudo-continuum. We provide both Li and H$\alpha$ EW measurements in Table~\ref{T_Temp_EW_Abun_LYS}.

\subsection{Radial velocities}\label{S_RV}

We used the \,{\sc fxcor} procedure in \,{\sc iraf}, which cross-correlates pixels between the target objects and those of template spectra to determine a relative RV \citep{1979a_Tonry}. SNR improves towards the red end of the spectrum, therefore measurements were made over 9 consecutive orders ($\lambda\lambda$ 5920--6520\AA), avoiding the broad Na D lines and contamination from telluric absorption. To determine the centre of the CCF peak a Gaussian fit was made to pixels within the top 40 per cent of the normalised peak. Some orders match better than others and poor CCF matches could be due to a spectral-type mismatch between an RV standard and a target, too much noise in the data, or, if the target is a close binary, then the CCF may contain multiple (unresolved) peaks. The spectrum of a binary companion could be blended and more apparent in some orders than others. In some cases, no CCF peak was observed at all. In order to provide a more robust RV measurement (and also assess the precision) the clipped mean RV from 9 orders. Any measurements $> 2\sigma$ from the mean were removed, and the process iterated a maximum of ten times. A weighted average relative RV for each target was calculated over each order for each template star using the $R$ quality factor in \cite{1979a_Tonry}\footnote{This is the ratio of the height of the true peak in the CCF compared to the average CCF value over the whole spectral order.}. In all cases the template that provided the lowest error in RV was used. Finally the RVs were transferred onto a heliocentric reference frame.

There were three sources of error: the standard error measured over the echellogram orders (weighted by $R$), the published RV uncertainty of the standard star used in the calibration and a systematic uncertainty estimated from the cross-correlations of all the RV standards with one another. These errors were treated as independent and were added in quadrature. RV measurements and errors are provided in Table~\ref{T_RVs_LYS}, along with any previous literature values. The RV measurements for each RV standard relative to one another formed a matrix of cross-correlation values. The systematic uncertainty was measured to be $0.36\,{\rm km\,s}^{-1}$ for the templates used in the June 2011 run and $0.27\,{\rm km\,s}^{-1}$ for the templates used in the December 2012 run. To avoid any spectral mismatch between target and template calibrations, templates are restricted to within 0.5 spectral classes of the target star in each case. RVs obtained from INT spectroscopy were measured using a similar procedure, but this time separating the long-slit spectra into ten 50\AA\ windows between 6600 and 7100\AA\ and obtaining the mean RV, weighted by $R$. These measurements were not included in the final averaged RV however because the overall accuracy was much poorer, limited by systematic uncertainties of $\sim 1\,{\rm km\,s}^{-1}$ in the wavelength calibration, were used only to support the data obtained from the NOT.

From the initial sample of 146 targets, 28 were found to have indeterminate RVs (as a result of poorly constrained cross-correlation peaks due to noisy data, high $v \sin i$ or perhaps binarity) and 14 objects were observed to have RVs that varied by more than $5\,{\rm km\,s}^{-1}$ on the timescale of the observing run, presumably as a result of binarity (see $\S$\ref{S_Bin_Con}).

\input{RVs_LYS}

\subsection{Projected rotational velocities}\label{S_vsini}

Projected rotational velocities ($v\sin i$) were estimated from the FWHM of the CCF using a set of simulations for objects observed on the NOT. Broadened versions of a number of slowly rotating template stars ($v\sin i$ templates, see Table~\ref{T_RV_Templates}) were produced by convolving high SNR spectra ($\simeq 100$) with a rotational broadening convolution kernel, using a limb-darkening parameter $\epsilon = 0.6$. Because the $v\sin i$ templates all had a small, but finite, rotational speed ($< 5\,{\rm km\,s}^{-1}$) a correction was made for the small measured $v\sin i$ values of the standards to set their non-rotating profiles at zero, and artificial Gaussian noise was injected into the spun-up spectra. For stars with large rotational broadening, the high-frequency (wavenumber $k > 400\,{\rm m}^{-1}$) components in the spectra were filtered out to minimise the effects of Gaussian noise (and the same filtering also applied to the target spectra). Each standard was broadened in $1\,{\rm km\,s}^{-1}$ steps between 5 and $70\,{\rm km\,s}^{-1}$ and the FWHM of each broadened standard cross-correlation peak was recorded using \,{\sc fxcor} and repeated for the same echellogram orders used in the RV analysis. Relationships between FWHM of the CCF and $v\sin i$ were derived using the means from the orders weighted by the $R$ quality factor (see \citealt{1979a_Tonry}) for each template. To ensure there were no spectral mismatches, only $v\sin i$ template spectra within 5 spectral sub-classes of the target were used in each case. The resolving power and SNR for spectra on the INT were insufficient to obtain $v\sin i$ measurements. The measured $v\sin i$ values are presented in Table~\ref{T_Kin_LYS}.

\subsection{Effective temperatures and Li abundance calculations}\label{S_Temp}

Effective temperatures were estimated based on a fifth order polynomial interpolation of the $V-K_{\rm s}$ and $T_{\rm eff}$ values between F0 and M3 stars provided in table 5 of \cite{2013a_Pecaut} and spectral-types are calculated from their linear interpolation with $V-K_{\rm s}$ in the same table. There is evidence to suggest that the metallicities of young, nearby stars do not vary much from the solar value \citep{2014a_Spina} and FGK stars should be close to the ZAMS by 20\,Myr. The EW of the Li\,{\sc i}~6708\AA\ line is transformed into logarithmic Li abundances (on the usual scale where $\log N$(H)$= 12$) using the curves of growth from \cite{1993b_Soderblom} and correcting for non-local thermodynamic equilibrium effects using the code provided in \cite{1994a_Carlsson}. Table~\ref{T_Temp_EW_Abun_LYS} displays the temperatures, EWs, Li abundances and their associated uncertainties.

\input{Temp_EW_Abun_LYS}

\subsection{Binary contaminants}\label{S_Bin_Con}

Tidally-locked, short-period binaries (herein, TLSPBs) maintain high rotational velocities well into their main sequence lifetimes and therefore can retain short rotation periods into old age. According to numerical calculations, solar-mass binaries starting from the birthline that have periods $\lesssim 8$ days will have circular orbits and synchronous rotation upon reaching the main sequence \citep{2005a_Stahler}. They will not spin down with age like single stars, but their rapid rotation will ensure they maintain their magnetic activity levels. Furthermore there is evidence that TLSPBs may deplete their lithium at a different rate to single stars (e.g. \citealt{2001a_Barrado_y_Navascues, 2011a_Canto_Martins, 2012a_Strassmeier}). For these reasons we attempt to identify these objects in our sample and when we find them we do not attempt to estimate their age. 

TLSPBs are expected to be the main contaminants in our survey, but should have large RV variations detectable by two observations separated by $\sim 24$ hours. In general, the RVs of likely-single, slow rotators with a reasonable SNR could be measured to a precision of $\sim 1\,{\rm km\,s}^{-1}$. We set a criterion that an RV difference $> 5\,{\rm km\,s}^{-1}$ between 2 measurements of the same object on separate nights identifies these objects as close binary systems. A numerical grading system is used to distinguish likely single stars (allocated a grade of 1) from those very likely to be binaries (allocated a grade of 5). A score of 5 was given if we detected RV differences $> 5\,{\rm km\,s}^{-1}$ for a target on separate nights or (if there was only one measurement) if there were literature sources indicating that either the object is a close binary or report a RV measurement $> 5\,{\rm km\,s}^{-1}$ discrepant from our measurement. In addition, to score 5, the average error in RV measurement must be less than $5\,{\rm km\,s}^{-1}$. Objects are graded 4 if there is only one spectrum which results in a) a clear, multi-peaked cross-correlation function (CCF) and b) an RV error $< 5\,{\rm km\,s}^{-1}$. A grade of 3 was given if the status of the star from the CCF was unclear (presumably as a result of poor SNR and/or large $v\sin i$), resulting in either an indeterminate RV or a RV uncertainty $> 5\,{\rm km\,s}^{-1}$. A grade of 2 was used for objects which had a single spectrum, an RV uncertainty $< 5\,{\rm km\,s}^{-1}$ and a distinct single peak in the CCF. Finally, objects scoring 1 had consistent, low-uncertainty RV measurements for 2 or more spectra. Individual RV measurements and binary scores for each object are presented in Table~\ref{T_RVs_LYS}.

From the entire observed sample of 146 objects, 14 scored 5, 29 of them scored either 4 or 3, 45 scored 2 and 29 objects scored 1. Any objects that scored 5 were automatically assumed to be close binaries. All targets selected for further analyses in $\S$\ref{S_Indiv} have binary scores of either 1 or 2. However, 2 or more consistent RV values for an object is not a guarantee that the object is not a TLSPB. Given that some of the objects in this sample have orbital periods on the order of $\sim 24$ hours, consistent RV measurements could be a result of measuring two points at the same phase on an RV curve.

\subsection{Binary simulation}\label{S_Bin_Sim}

Although the criterion that two separate observations resulting in RV differences $> 5\,{\rm km\,s}^{-1}$ is sufficient to flag these objects as TLSPBs, a fraction with RV differences $< 5\,{\rm km\,s}^{-1}$ may remain undetected. In a TLSPB system the rotation period will be equal to the orbital period. Given that these observations are incapable of distinguishing between TLSPBs and single stars, a simulation was carried out to calculate the probability that each object showing no RV variations (i.e. not $> 5\,{\rm km\,s}^{-1}$) was in fact a short period binary seen at low inclination or at two similar phases in its orbit. 

The simulation used $10^{6}$ iterations, a flat mass-ratio between 0.20 and 0.95 \citep{1990a_Hogeveen} and the primary mass was estimated from its spectral-type. Tidally-locked binaries will be circular, therefore the eccentricity is zero in all calculations. A value between 0 and $2\pi$ was randomly assigned for the argument of periastron. A second RV measurement was calculated at a later time equal to the cadence (cad) between observations.

\begin{figure}
    \begin{center}
    \vspace{0.00mm}
     \begin{minipage}{1.0\linewidth}
            \centering
            \includegraphics[width=1.0\textwidth]{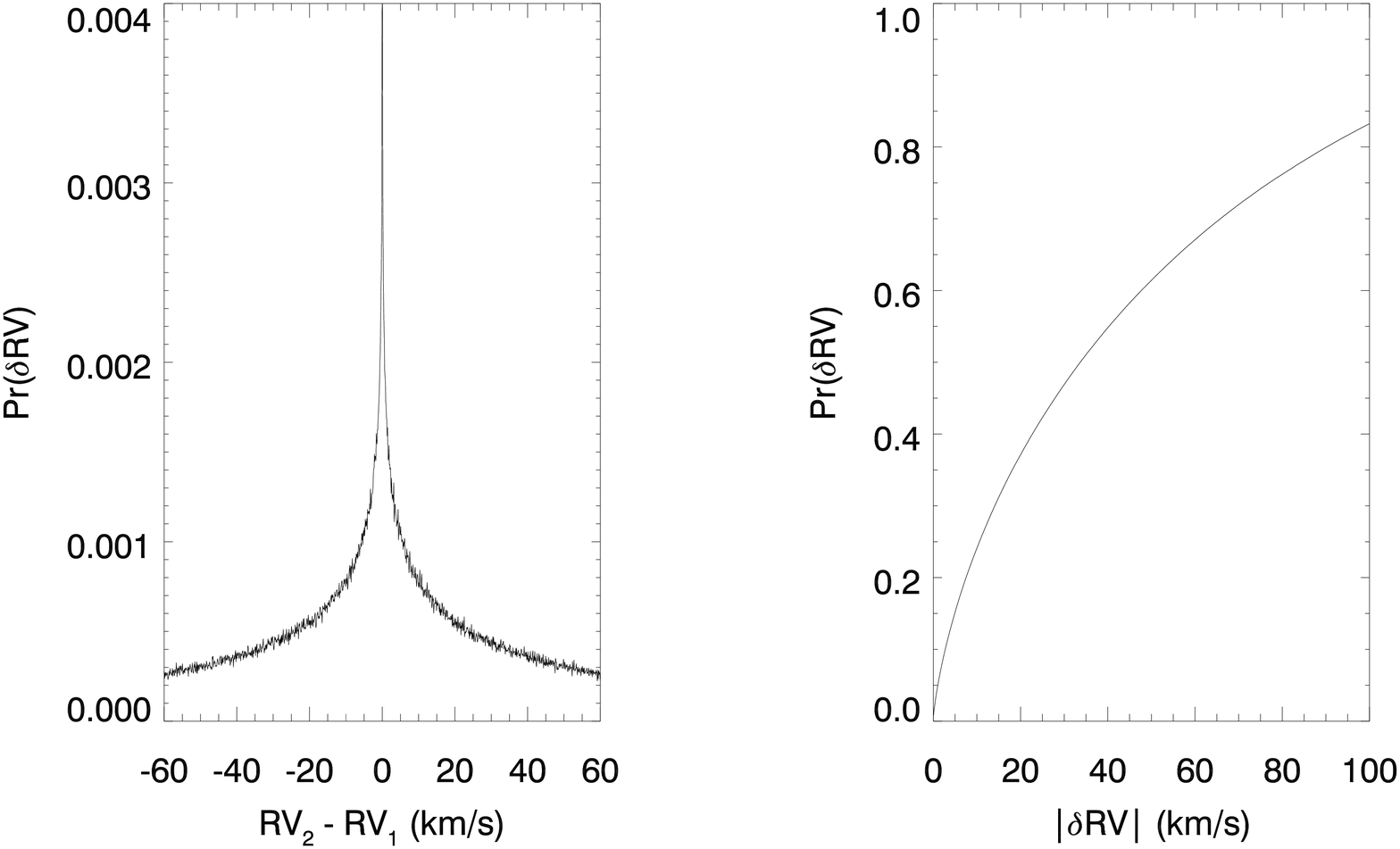}
         \end{minipage}
    \vspace{0.00mm}
        \begin{minipage}{1.0\linewidth}
           \centering
    \vspace{10.00mm}
            \includegraphics[width=1.0\textwidth]{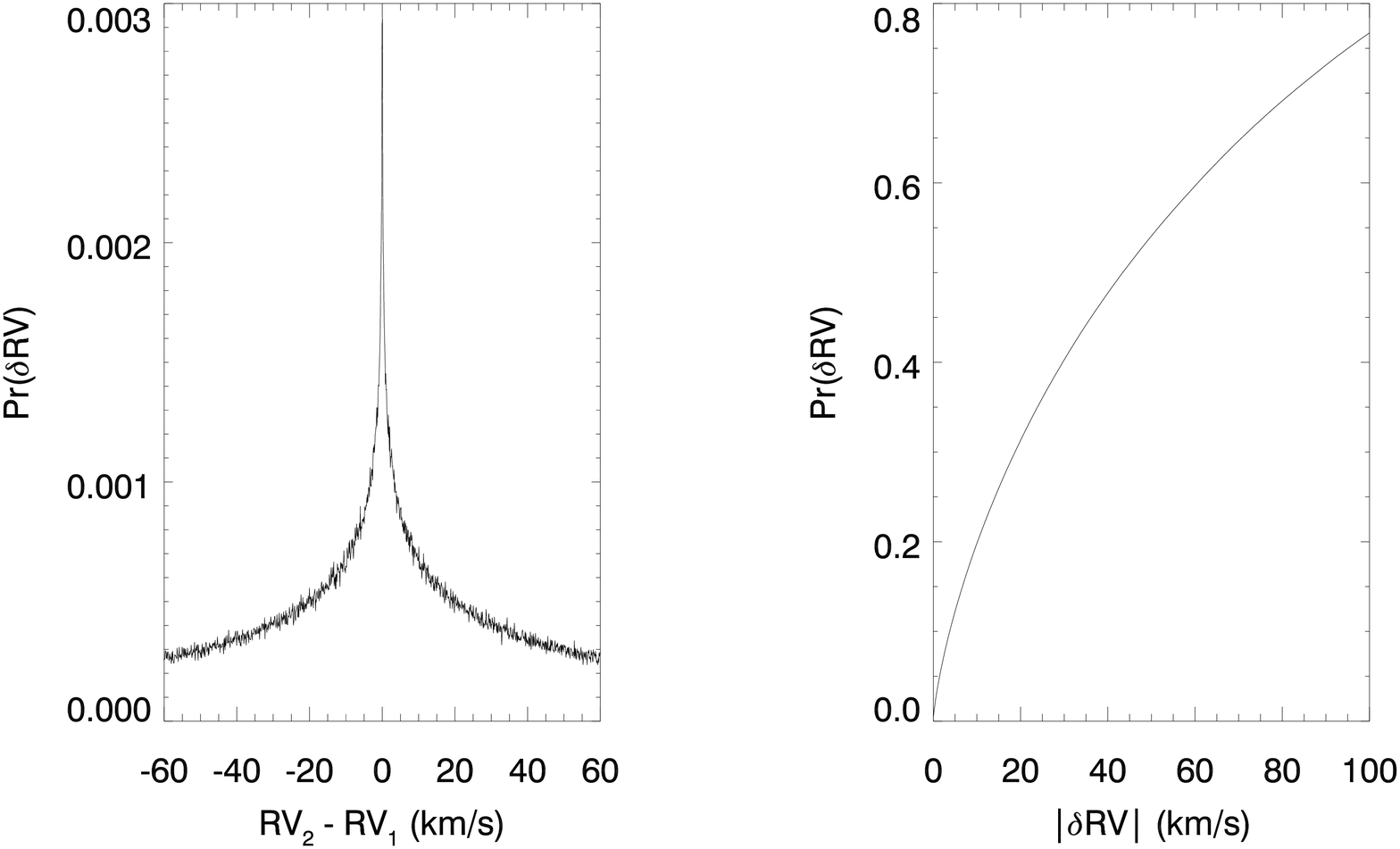}
         \end{minipage}
    \hspace{0.00mm}
    \end{center}
        \begin{flushleft}
 \caption{Simulations to calculate the fraction of tidally-locked binary objects that would result in an RV difference $< 5\,{\rm km\,s}^{-1}$ over the cadence of two observations for target D (top) and target P (bottom). The input parameters and a description of the simulation is provided in $\S$\ref{S_Bin_Sim}. The left panel demonstrates the probability that a binary star has a particular RV difference of a certain value. The right panel is the cumulative probability that an object has $|{\rm RV}_{2}-{\rm RV}_{1}|$ less than a given value.}
  \label{F_Bin_Sim}
    \end{flushleft}
    \end{figure}

We carried out the simulation for all objects that had 2 separate RV measurements at the NOT. In some cases, the separation between the observations was similar to an integer number of rotation periods. In these cases, a TLSPB nature cannot be ruled out with any confidence but for the majority of cases we are $\sim 90$ per cent confident that they cannot be TLSPBs. In other words in $> 90$ per cent of the simulations there would have been a $\Delta{\rm RV} > 5\,{\rm km\,s}^{-1}$. Simulations for two objects, target P ($P = 1.04$\,days, ${\rm cad} = 2.00$\,days, the probability that the object is a TLSPB ($P_{\rm TLSPB}$) = 0.12) and target D ($P = 3.85$\,days, ${\rm cad} = 0.97$\,days, $P_{\rm TLSPB} = 0.15$) are presented in Figure~\ref{F_Bin_Sim}. In the initial target selection, light-curves indicative of eclipsing binaries were filtered out, therefore it is unlikely that high inclination, eclipsing TLSPBs have been included. For example, a 0.7$M_{\odot}$/0.5$M_{\odot}$ binary system with an orbital period of 3 days would be at least partially eclipsing if the inclination angle, $i > 60^{\circ}$. Given a random orientation of orbital rotation axes, the probability of observing an object with $i > 60^{\circ}$ is 0.5, therefore $P_{\rm TLSPB}$ could be over-estimated by a factor of 2.

The incidence of close binaries in a random sample of field stars is expected to be quite low. We simulated a field population based on the binary fraction, period and eccentricity distribution of field solar-type stars proposed by \cite{2010a_Raghavan}, finding that only 2.2 per cent of such a sample should lead to observed RV differences of $>5\,{\rm km\,s}^{-1}$ in observations separated by $\sim 1$ day. However, of the 55 objects that we observed twice, 11 were found to show evidence of close binarity -- significantly higher than 2.2 per cent. The reason is likely that we are not observing random field stars, but a sample of fast-rotating, active stars, which will tend to be either young or in close binary systems (e.g. \citealt{1995a_Jeffries}). 

The field star simulation served an additional purpose. Where we identify a Li-rich, potentially young star, with two consistent RV measurements (within $5\,{\rm km\,s}^{-1}$), then assuming that these stars share the same binary properties as average field stars, there is still a 7.2 per cent probability that the true systemic velocity could be different to that measured by $> 5\,{\rm km\,s}^{-1}$ due to the presence of an unseen or unresolved wide binary companion.

\section{Finding young stars}\label{S_Age}

\subsection{Gyrochronology}\label{S_Gyro}

Gyrochronology provides a colour-dependent method to obtain an age estimate for single, late-type stars by using empirical fits to rotation rates in open clusters of known age. Figure~\ref{F_Gyrochrones} displays the rotation periods against $V-K_{\rm s}$ for the entire observed sample, overplotted with curves for the interface (I) and convective (C) sequences of the form defined by in Barnes (2003; 2007). \cite{2008a_Mamajek} find age discrepancies between the calibrated gyrochrones in Barnes (2003; 2007), claiming that the Barnes calibrations would underestimate the Hyades age. We therefore use the updated calibration in \cite{2008a_Mamajek} and transform $B-V$ to $V-K_{\rm s}$ by interpolating table 5 in \cite{2013a_Pecaut}. At this point we stress two issues about our use of gyrochronology; firstly, the calibrations in \cite{2008a_Mamajek} are for stars with $0.5 \leq (B-V)_{0} \leq 0.9$ ($\sim$ F7--K2) and roughly half of our sample are redward of this range. Secondly, we find that in almost all of our sample the upper age limits are not generally affected by the choice of either the \cite{2007a_Barnes} or \cite{2008a_Mamajek} calibrations.

The bifurcation point is the point at which the C sequence meets the I sequence and the isochrones become bimodal redward of this point. To estimate an age using gyrochronology the following analysis is carried out (see Figure~\ref{F_Gyrochrones}):

\textbullet If the object lies above the I sequence, it is definitely older than this gyrochrone.

\textbullet If the object lie below the I sequence and blueward of the
   the corresponding C sequence, then this target is younger than that gyrochrone.

\textbullet If the object lies below the I sequence and redward of
   the corresponding C sequence, there is an ambiguity and one must look for an
   older gyrochrone which satisfies the previous condition.

To clarify how gyrochronology works in practice we highlight three stars from Table~\ref{T_PX_LYS} (targets A, B and L):

\noindent\textbf{Target A} - Target A lies below the 100\,Myr I sequence, but above the 50\,Myr I sequence. It is redward of the 10, 50 and 100\,Myr C gyrochrones, but lies blueward of the 300\,Myr C gyrochrone and the 300\,Myr bifurcation point. Hence the age of target A is estimated to be $< 300$\,Myr, with a lower limit of 50\,Myr. 

\noindent\textbf{Target B} - Target B has a minimum age of 50\,Myr, as it lies above the I sequence of this isochrone. It is below the I sequence of all gyrochrones with ages $\geq$ 100\,Myr but is only blueward of the C sequence for those with ages $\geq 100$\,Myr. Hence the age of B is estimated to be between 50 and 100\,Myr. 

\noindent\textbf{Target L} - Target L is at least older than 100\,Myr, because it is above the I sequence at this age. Although it lies below all the I sequences $\geq 100$\,Myr it is only blueward of the C sequence at an age of 300\,Myr. Hence L has a gyrochronology age between 100 and 300\,Myr.

The limitations of the initial target selection in terms of colour and period now become clearer when examining Figure~\ref{F_Gyrochrones}. A period $< 5$ days implies an age $< 700$\,Myr for F-stars; $< 300$\,Myr for G-stars with $1.5 < V-K_{\rm s} < 2.0$, but then increases again to $< 700$\,Myr for K-stars with $2.0 < V-K_{\rm s} < 3.0$ and even older for M-dwarfs. Reversing this argument, to be sure an object has an age $\leq 100$\,Myr would require it to have a period of $< 1$ day if it were a late F-star, $< 2$ days for a G-star and then less than 0.7 of a day for K-stars and less than 0.3 of a day for M-stars. To avoid ambiguity with interpreting the C-sequence we provide (and employ) only upper-age limits from gyrochronology for the likely-young sample and these are provided in Table~\ref{T_Ages_LYS}.

Gyrochronology will not work for tidally-locked binary systems. Because this is the likely status of all the objects flagged as 4 or 5 in the binarity tests, we choose not to assign ages to them based on gyrochronology (or activity - see section~\ref{S_Ha_EW}).

\subsection{Li EW measurements compared to known clusters}\label{S_Li_EW}

The strength of photospheric lithium in cool stellar atmospheres can be used as an empirical age indicator. Stars are born with the Li abundance of the interstellar medium, but it is depleted by proton capture in their interiors. Mixing processes bring the Li-depleted material to the surface, resulting in an age- and mass-dependent photospheric Li-abundance (see \citealt{2005a_Sestito, 2006a_Jeffries}). 

Whilst theoretical models partially capture the behaviour of PMS Li depletion in FGK stars, they are strongly sensitive to the assumed opacities in the atmospheres and interiors of these stars and cannot readily explain the spread in Li abundance that is seen at a given effective temperature in presumably coeval clusters. For example, the spread in Li abundance amongst K-stars in the Pleiades is $1-2$\,dex \citep{1993a_Soderblom}. For these reasons Li is used here only as an empirical age estimator by comparing the Li EWs of targets, as a function of $B-V$ and $V-K_{\rm s}$ (or equivalently, Li abundance versus $T_{\rm eff}$), with stars observed in the Hyades (age $\sim 625$\,Myr, \citealt{1998a_Perryman}) and Pleiades (age $\sim 125$\,Myr, \citealt{1998a_Stauffer}) clusters. Li EW data obtained from Soderblom et al. (1990,1993a,b,c,1995a,b); Jones et al. (1996,1997); Jones, Fischer $\&$ Soderblom (1999); Soderblom et al. 1999 and \cite{2002a_Wilden}. In addition, comparisons were made with three younger clusters; NGC~2264 (age $\sim 5$\,Myr, \citealt{1998a_King, 1999a_Soderblom, 2005a_Dahm}), $\gamma$~Vel (age $\sim 10$\,Myr, \citealt{2014b_Jeffries}) and IC~2602 ($\sim 30$\,Myr, \citealt{2000a_Meola, 2001a_Randich, 2001b_Randich}).

\nocite{1990a_Soderblom}
\nocite{1993a_Soderblom}
\nocite{1993b_Soderblom}
\nocite{1993c_Soderblom}
\nocite{1995a_Soderblom}
\nocite{1995b_Soderblom}
\nocite{1996a_Jones}
\nocite{1997a_Jones}
\nocite{1999a_Jones}
\nocite{1999a_Soderblom}

    \begin{figure}
    \begin{center}
    \vspace{0.00mm}
     \begin{minipage}{1.0\linewidth}
            \centering
            \includegraphics[width=1.0\textwidth]{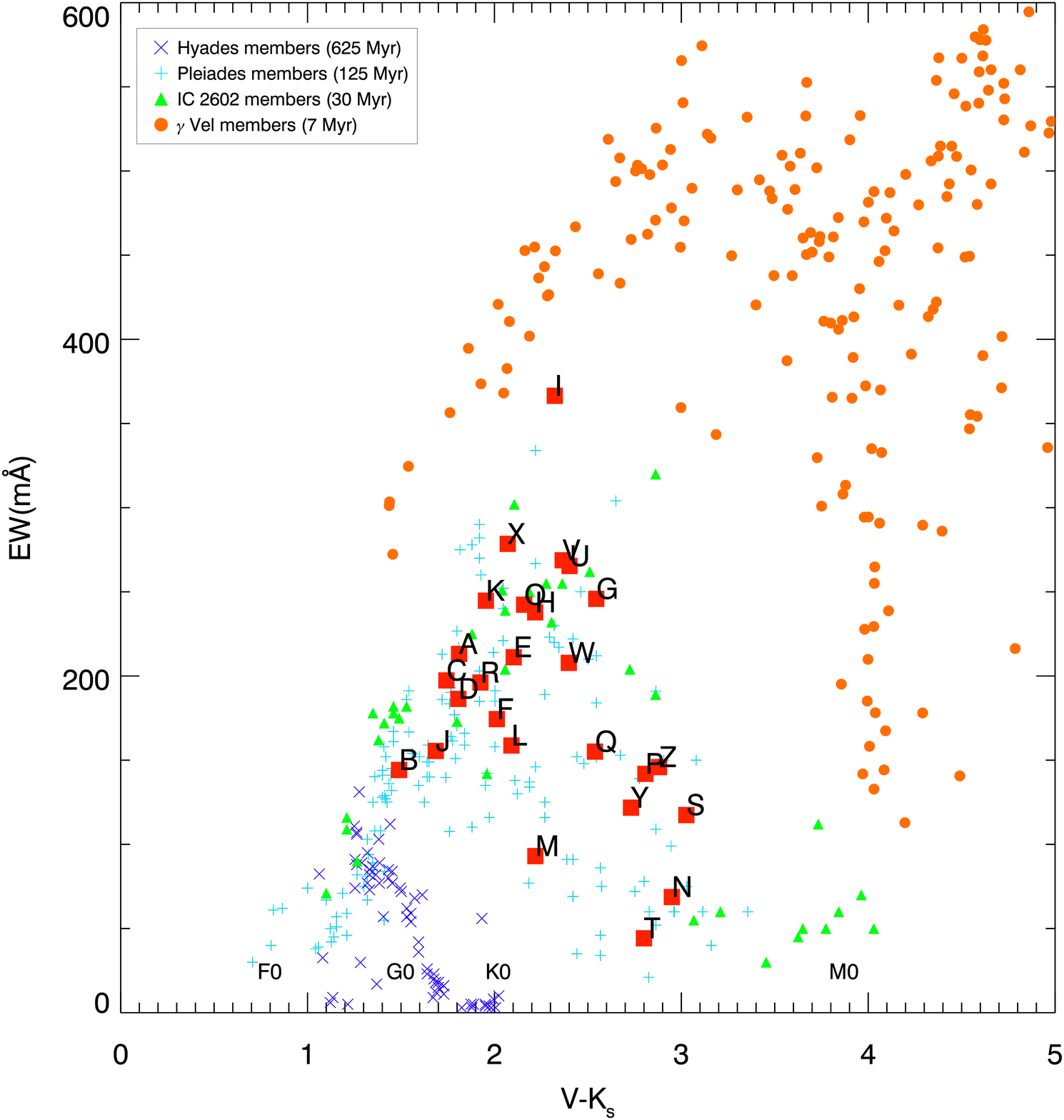}
         \end{minipage}
    \vspace{0.00mm}
        \begin{minipage}{1.0\linewidth}
           \centering
    \vspace{1.0cm}
            \includegraphics[width=1.0\textwidth]{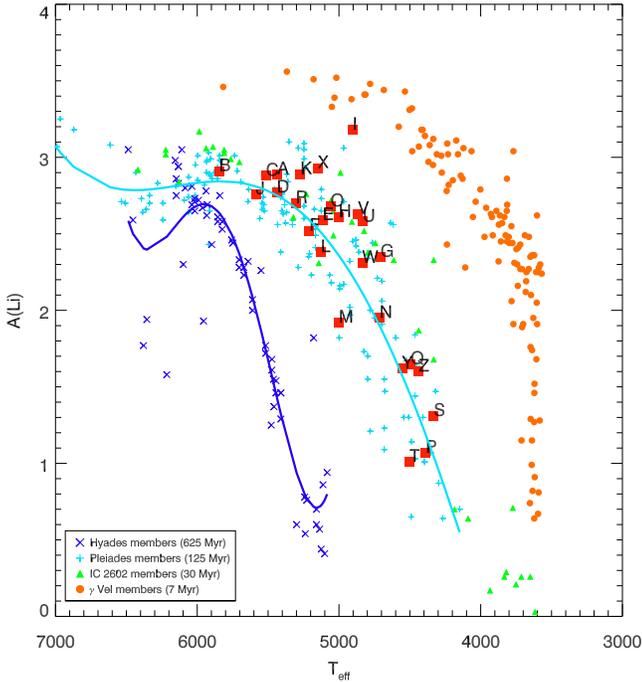}
         \end{minipage}
    \end{center}
        \begin{flushleft}
 \caption{Top: Li EWs and corresponding $V-K_{\rm s}$ colours. Green open circles and filled diamonds represent the same objects described in Figure~\ref{F_Gyrochrones}. Objects are compared to members in the Hyades (625\,Myr), the Pleiades (125\,Myr), IC~2602 (30\,Myr) and $\gamma$~Vel (10\,Myr, see text for references) to estimate an Li-based age range. Error bars are not included, but are provided in Table~\ref{T_Temp_EW_Abun_LYS}. Bottom: Li abundances as a function of surface temperature. Both the target sample and all ancillary data from the open clusters are folded through the same curve of growth (described in $\S$\ref{S_Temp}). Fourth-order polynomial fits are overplotted for the Hyades and Pleiades, however these are not implied to represent the trend of Li abundances in these clusters. We display only the objects from Table~\ref{T_PX_LYS} in the bottom panel and Li abundances for all other objects are provided in the online supplementary data.}
      \label{F_Li_EW}
    \end{flushleft}
    \end{figure}

In Figure~\ref{F_Li_EW} we present the Li EWs and colours for the entire observed sample, and ages are estimated using comparisons to the Li patterns in the aforementioned open clusters. One object (labelled as target I) with duplicate, consistent RV measurements has a Li EW/$V-K_{\rm s}$ indicative of stars younger than the Pleiades (more similar to the pattern observed in $\gamma$~Vel at $\sim 10$\,Myr). A further twenty-five targets with binary scores of 1 or 2 have Li EWs/colours (or Li abundances/temperatures) consistent with the Pleiades or IC~2602. The remainder of the targets appear to have depleted their Li sufficiently that they appear older than the Pleiades, or they are close binaries and their Li-depletion may not properly reflect their true age. There is some evidence that close binary system components have anomalous Li-depletion histories (see, for example \citealt{2012a_Strassmeier}). The bottom plot of Figure~\ref{F_Li_EW} is A(Li) versus $T_{\rm eff}$. As this is essentially the data from the top two plots of Figure~\ref{F_Li_EW} folded through a transformation, it contains no new information on the ages, but does illustrate that the most Li-rich stars found are still depleted from an assumed initial level of about A(Li)=3.3, so are probably not very young PMS stars. The estimated Li age ranges for the entire observed sample are presented in Table~\ref{T_Ages_LYS}.

\subsection{X-ray activity in the ROSAT catalog}\label{S_Xray}

The initial selected targets described in $\S$\ref{S_Obs} were chosen based on their short rotation periods and assessment of apparent rotation modulation in their light-curves. All such objects originated from a catalog of objects that had entries in both ROSAT and SuperWASP. Throughout the selection process no criteria were used based on any specific X-ray property -- a database entry in either the 1RXS or 2RXP catalog was deemed sufficient. Data were available for all the selected targets and were analysed subsequent to the telescope observations. $L_{\rm X}/L_{\rm bol}$ was calculated for each object using the following formula:

\begin{equation}
  \frac{L_{\rm X}}{L_{\rm bol}} = \frac{f_{\rm X}}{2.48 \times 10^{-5} \times 10^{-0.4m_{\rm bol}}}
  \label{E_LX}
\end{equation}

The ROSAT catalogs were revisited to obtain count rates, exposure times and hardness ratios (CR, exp and HR1, respectively, all provided in Table~\ref{T_PX_LYS}). X-ray fluxes ($f_{\rm X}$) were calculated by multiplying the count rate by the energy conversion factor (ECF) provided in \cite{2001a_Stelzer}:

\begin{equation}
   {\rm ECF} = (8.31 + 5.30{\rm HR1}) \times 10^{-12} {\rm erg\,cm^{-2}\,cr^{-1}},
  \label{E_ECF}
\end{equation}

where ${\rm cr}^{-1}$ is the number of counts per second. Bolometric magnitudes were found using $V$ magnitudes and a main-sequence bolometric correction interpolated from $V-K_{\rm s}$ using table 5 in \cite{2013a_Pecaut}. Main-sequence conversions should be appropriate for the F, G and K stars in this sample, however, if stars were younger than 30\,Myr this may result in a small calibration error of $\sim 0.1$\,dex.

    \begin{figure}
    \begin{center}
            \includegraphics[width=0.5\textwidth]{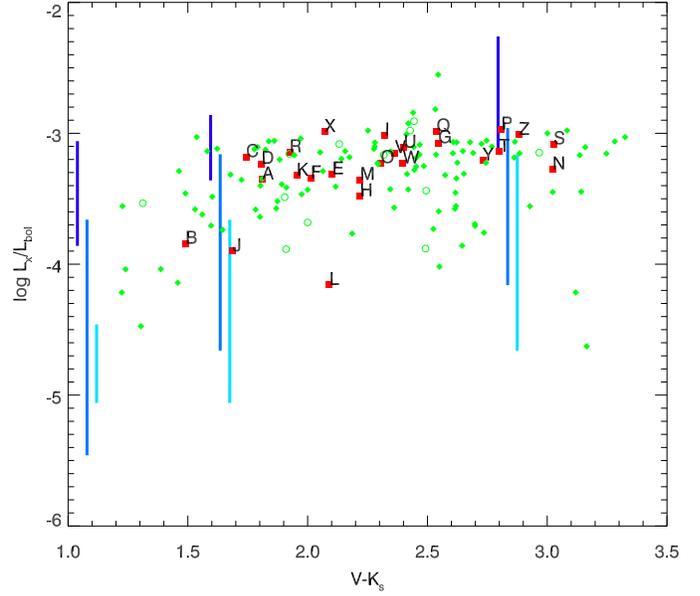}
    \end{center}
    \begin{flushleft}
\caption{$\log L_{\rm X}/L_{\rm bol}$ as a function of $V-K_{\rm s}$. The $10^{\rm th}$ to $90^{\rm th}$ percentile range of $L_{\rm X}/L_{\rm bol}$ values in NGC~2547, the Pleiades and the Hyades are represented by the light blue/blue/dark blue lines, respectively.}
 \label{F_LX}
\end{flushleft}
\end{figure}

Figure~\ref{F_LX} shows $L_{\rm X}/L_{\rm bol}$ as a function of $V-K_{\rm s}$. For reference, the $10^{\rm th}$ and $90^{\rm th}$ percentile $L_{\rm X}/L_{\rm bol}$ range for F, G and K stars in NGC~2547, the Pleiades and the Hyades (based on figure 12 in \citealt{2006b_Jeffries}) are represented by light blue/blue/dark blue lines (respectively). $L_{\rm X}/L_{\rm bol}$ is a relatively crude age indicator and largely rotation-dependent in any case. Almost all of the candidates could be consistent with a Pleiades age (as expected for their rotation rates). As it turned out the probable young objects were slightly more active than the average star in our spectroscopic sample.

\subsection{H$\alpha$ emission}\label{S_Ha_EW}

A potentially more direct activity-age dependent probe is to observe the strength of the H$\alpha$ line at 6563\AA. H$\alpha$ emission lines are diagnostic of strong magnetic activity in the photospheres of stars, which are linked to rotation and young ages. 

The empirical behaviour is that young stars show H$\alpha$ emission, but this emission is not apparent in stars with photospheres warmer than a temperature that appears to be age-dependent (\citealt{1995a_Reid, 1999a_Hawley}). Using data from clusters ranging from 30--625\,Myr (IC~2602, IC~2391, NGC~2516, Pleiades and Hyades) \cite{1999a_Hawley} derived a log-linear fit to the age (in Myr) of a cluster and the $V-I$ colour at which stars first display H$\alpha$ emission given as:

\begin{equation}
 (V-I)_{\rm emission} = -6.42 + 0.99(\log{\rm age/Myr})
\label{E_Ha_Emi}
\end{equation}

At redder colours there is a point where no stars have H$\alpha$ in absorption. Compiling data in 7 open clusters aged between 30 and 625\,Myr (NGC~2391 and NGC~2602, \citealt{1997a_Stauffer}; NGC~2547, \citealt{2000a_Jeffries}; Blanco~1, \citealt{1997a_Panagi}; Pleiades, \citealt{1993e_Soderblom}; NGC~2516, \citealt{1999a_Hawley}; Hyades, \citealt{1991a_Stauffer}) we derived a relationship for a given $V-I$ and age at which a star has reduced in magnetic activity and starts to show H$\alpha$ in absorption:

\begin{equation}
 (V-I)_{\rm absorption} = -4.19 + 0.68(\log{\rm age/Myr})\
\label{E_Ha_Abs}
\end{equation}

\begin{figure}
 \vspace{2pt}
 \begin{center}
\includegraphics[width=0.5\textwidth]{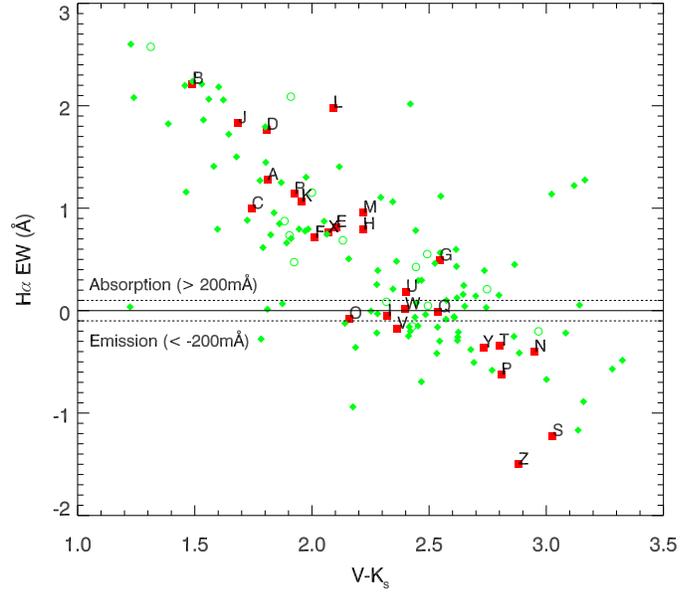}
 \end{center}
     \begin{flushleft}
 \caption{H$\alpha$ EW as a function of $V-K_{\rm s}$. Objects displaying emission may be compared to an upper age limit based on chromospheric activity and a lower age limit for H$\alpha$ absorption. Only objects with EWs $>$ 200\,m\AA\ ($< -200$\,m\AA) were identified as true emission (or absorption) lines.}
      \label{F_Ha_EW}
    \end{flushleft}
\end{figure}

These relationships can be used to estimate the minimum/maximum age of a star with an H$\alpha$ absorption/emission line. Targets with H$\alpha$ EW $> 200$\,m\AA\ are considered as emission lines (providing a maximum age) and EW $< -200$\,m\AA\ for H$\alpha$ in absorption (a minimum age). Targets with H$\alpha$ EWs in the range $\pm$ 200\,m\AA\ were considered as `filled-in' lines and equations~\ref{E_Ha_Emi} and~\ref{E_Ha_Abs} were used to provide a likely age \textit{range}. In Table~\ref{T_Temp_EW_Abun_LYS} the H$\alpha$ EWs for the likely-young sample are presented, and the ages estimated for the entire sample are provided in Table~\ref{T_Ages_LYS}. $V-I_{\rm C}$ colours are converted to $V-K_{\rm s}$ using table 5 in \cite{2013a_Pecaut}. The H$\alpha$ EWs of the entire observed sample are plotted as a function of their $V-K_{\rm s}$ baseline in Figure~\ref{F_Ha_EW}.

\subsection{Radii-colour isochrones}\label{S_Rsini}

\begin{figure}
 \vspace{2pt}
 \begin{center}
\includegraphics[width=0.5\textwidth]{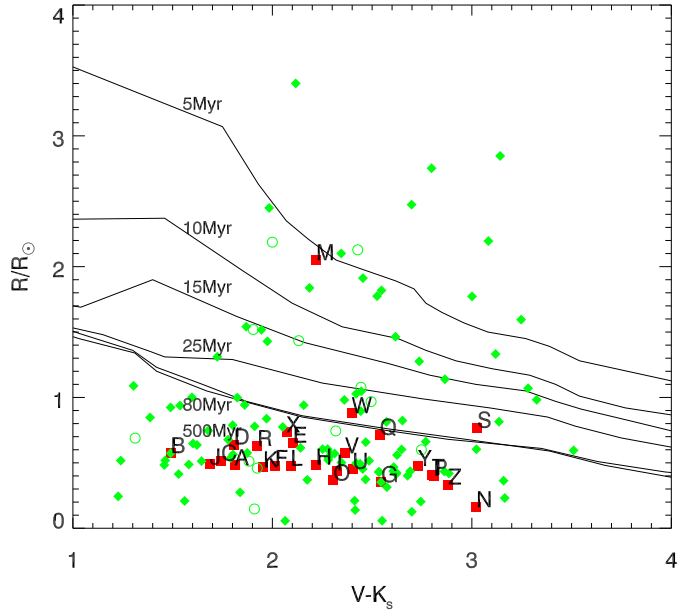}
\end{center}
    \begin{flushleft}
 \caption{Radii-colour relationship for the entire observed sample (minus the six objects with an indeterminate $v\sin i$ measurement). Continuous lines represent isochrones generated from the models in \protect\cite{2000a_Siess}. The plotted points are $R\sin i$ values and therefore lower limits to the true radius.}
      \label{F_R_VK}
    \end{flushleft}
\end{figure}

We calculated projected rotational velocities ($v\sin i$) following the procedures in $\S$\ref{S_vsini} and multiplied them by their rotational periods to obtain measurements of the projected stellar radius $R\sin i = 0.02\,Pv\sin i$ (in solar radii). Because of the $\sin i$ ambiguity, this technique is only capable of providing lower limits to the radius. Values of $R\sin i$ are compared to radius/$V-K_{\rm s}$ isochrones from the models of \cite{2000a_Siess}. Between ages of $10-100$\,Myr a star is contracting on the PMS, and because the radii calculated from $R\sin i$ are minima, this sets an upper limit to the age. The $R\sin i$ values used correspond to the 1$\sigma$ lower limit based on both the period and $v\sin i$ measurement, which sets a likely upper age limit. $R\sin i$ offers good age discrimination when $dR/dt$ is large, but becomes poor when $dR/dt \sim 0$ as objects approach the ZAMS. The targets that displayed no Li yet appear to have large $R\sin i$ values are more likely to be binaries as opposed to being genuinely young. One object in Figure~\ref{F_R_VK}, target M, has an $R\sin i$ age ($5-10$\,Myr) in contradiction with the Li age ($30-200$\,Myr) and is at odds with a H$\alpha$ lower limit of 32\,Myr.

\subsection{Combined age estimation and identification of young, single stars}\label{S_Age_Final}

In terms of providing a secure identification of youth, we regard \textit{Li as the primary age indicator}. Alternative age indicators are used only to support an age determination. If no Li is present in the spectra, then the object is automatically assumed to be at least older than the Hyades and is assigned a nominal age of 1\,Gyr (for the purpose of a parallax estimation -- see $\S$\ref{S_Parallax}).

Subsequent analysis in this paper is restricted to 26 objects that have Li-derived ages younger than 200\,Myr and have binary scores of 1 or 2. We categorise these objects as the \textit{likely-young sample}. Table~\ref{T_Ages_LYS} shows that, with the exception of target I, all of the likely-young sample have Li-based ages which overlap with the ages constrained from H$\alpha$. Each age estimate for the likely-young sample is provided in Table~\ref{T_Ages_LYS} (all other objects are listed online). $R\sin i$ values are only identified as younger than main-sequence for 3 objects, two of which (targets S and W) are consistent with Li and H$\alpha$ and the other, target M, is discussed in $\S$\ref{S_Rsini}. Target I has an age range of 5--30\,Myr, significantly younger than the rest of the sample.

\input{Ages_LYS}

\section{Kinematic analysis}\label{S_Kin}

\subsection{Parallax estimation}\label{S_Parallax}

Only four objects from the entire observed sample have a previously measured trigonometric parallax, only one of which is part of the likely-young sample. For the rest of the objects, parallaxes were estimated using $V-K_{\rm s}$ (assuming no reddening) and a maximum and minimum age (see Table~\ref{T_Ages_LYS}). Using the \cite{2000a_Siess} evolutionary models (with solar metallicity and no convective overshoot) a maximum and minimum absolute $K$ magnitude was calculated by interpolating $M_{K}$ for the youngest and oldest isochrone corresponding to the estimated age range, from which a `photometric' parallax was calculated.

\begin{figure}
 \begin{center}
\includegraphics[width=0.5\textwidth]{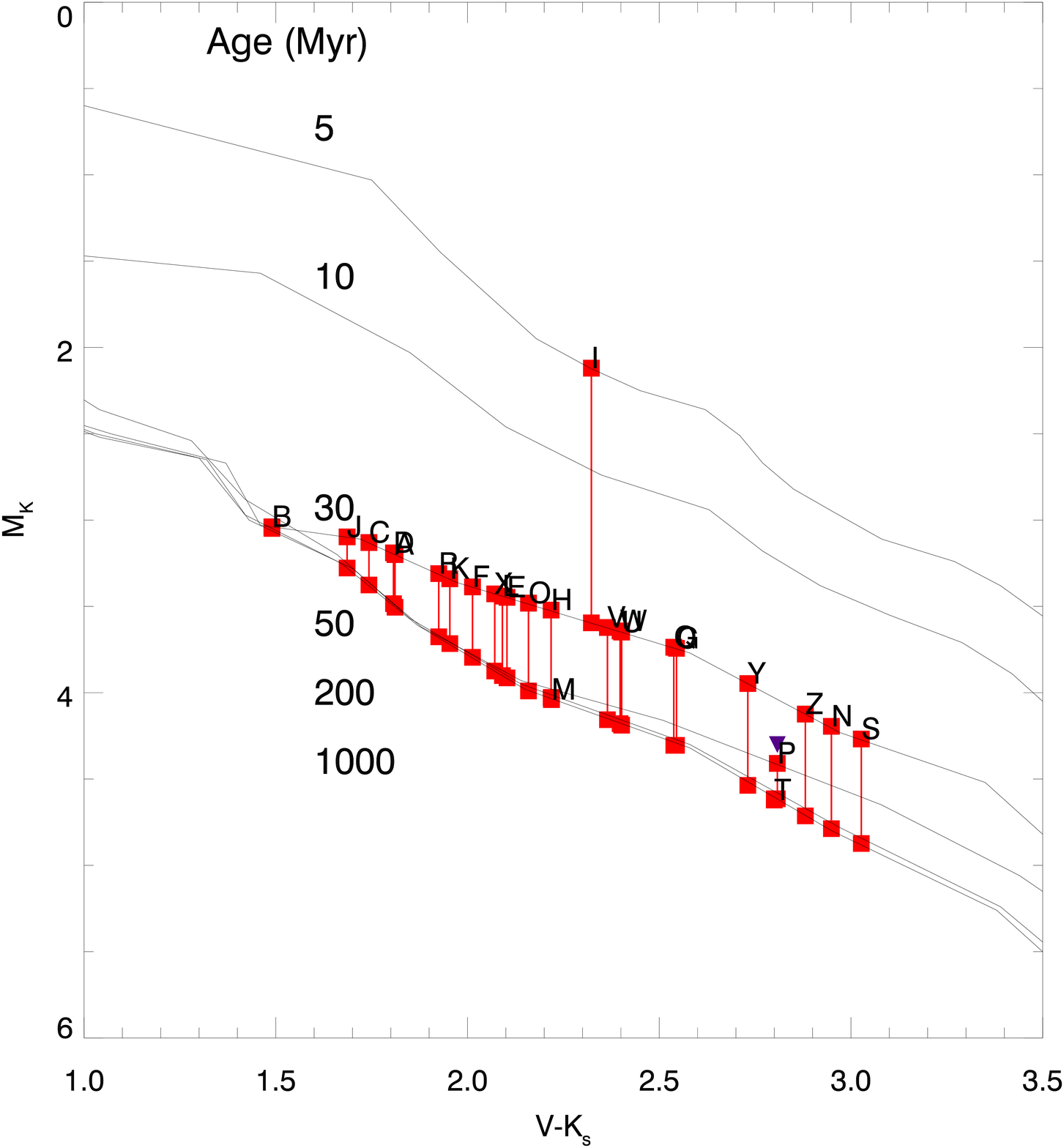}
\end{center}
    \begin{flushleft}
 \caption{$M_{K}$ vs $V-K_{\rm s}$ CMD for the entire observed sample. Red lines indicate the range in $M_{K}$ for the likely-young sample based on the assumed age range for each object. The model isochrones of \protect\cite{2000a_Siess} are overplotted at 5, 10, 30, 200 and 1000\,Myr. The purple downwards-facing triangle is the absolute $K$ magnitude for target P, measured from a trigonometric parallax.}
      \label{F_CMD}
    \end{flushleft}
\end{figure}

In Figure~\ref{F_CMD} the $M_{K}$ range is plotted for the likely-young sample, along with all other objects and TLSPBs. The plot shows that the younger objects span a larger range in $M_{K}$ due to PMS contraction. For example, target I, with an age range from 5 to 30\,Myr has a possible $M_{K}$ range of $\sim 1.5$\,mag, whereas target V, with an age ranging between 30 and 70\,Myr results in a difference in $M_{K}$ of $\sim 0.5$\,mag. Redward of $V-K_{\rm s} = 3.0$ isochrones are separated by no more than 0.7\,mag between 30\,Myr and the ZAMS and no more than 0.3\,mag between 50\,Myr and the ZAMS.

To test the precision of the photometric parallaxes, absolute $K$ magnitudes for the 4 objects with trigonometric parallaxes were estimated in the same way; one of which, target P, is a member of the likely-young sample, and is plotted as a purple downwards-pointing triangle. For each measurement the photometric parallax appears to under-predict the trigonometric parallaxes. The average difference between the $M_{K}$ measured from trigonometric parallax and estimated $M_{K}$ is $\sim 0.4\,$mag and the average range in $M_{K}$ from photometric parallaxes is 0.44\,mag. The parallax range for all objects is provided in column 6 of Table~\ref{T_Kin_LYS}.

\subsection{Calculating galactic space velocities}\label{S_UVW}

Using the prescription in \cite{1987a_Johnson}, Galactic space velocities (and their errors) were computed in terms of $U, V$ and $W$ (where $U$ points towards the Galactic centre, $V$ in the direction of Galactic rotation and $W$ in the direction of the Galactic north pole). The photometric parallax range is used if no trigonometric parallax is available. Table~\ref{T_Kin_LYS} lists all of the required input parameters and columns 5, 6 and 7 are the calculated $UVW$. Proper motions are from the PPMXL catalog (\citealt{2010a_Roeser}). Two separate error bars are quoted, the first includes contributions from the $\sigma_{\mu_{\alpha}}$, $\sigma_{\mu_{\delta}}$ and $\sigma_{\rm RV}$ uncertainties and the second corresponds to half of the range in each velocity coordinate resulting from the extrema of the possible photometric parallaxes (provided in column 6). Should a trigonometric parallax exist, or if an object is assigned a single-valued main-sequence age (1\,Gyr) then all errors are incorporated into one single error bar.

\begin{figure}
    \begin{center}
    \vspace{0.00mm}
     \begin{minipage}{1.0\linewidth}
            \centering
            \includegraphics[width=1.0\textwidth]{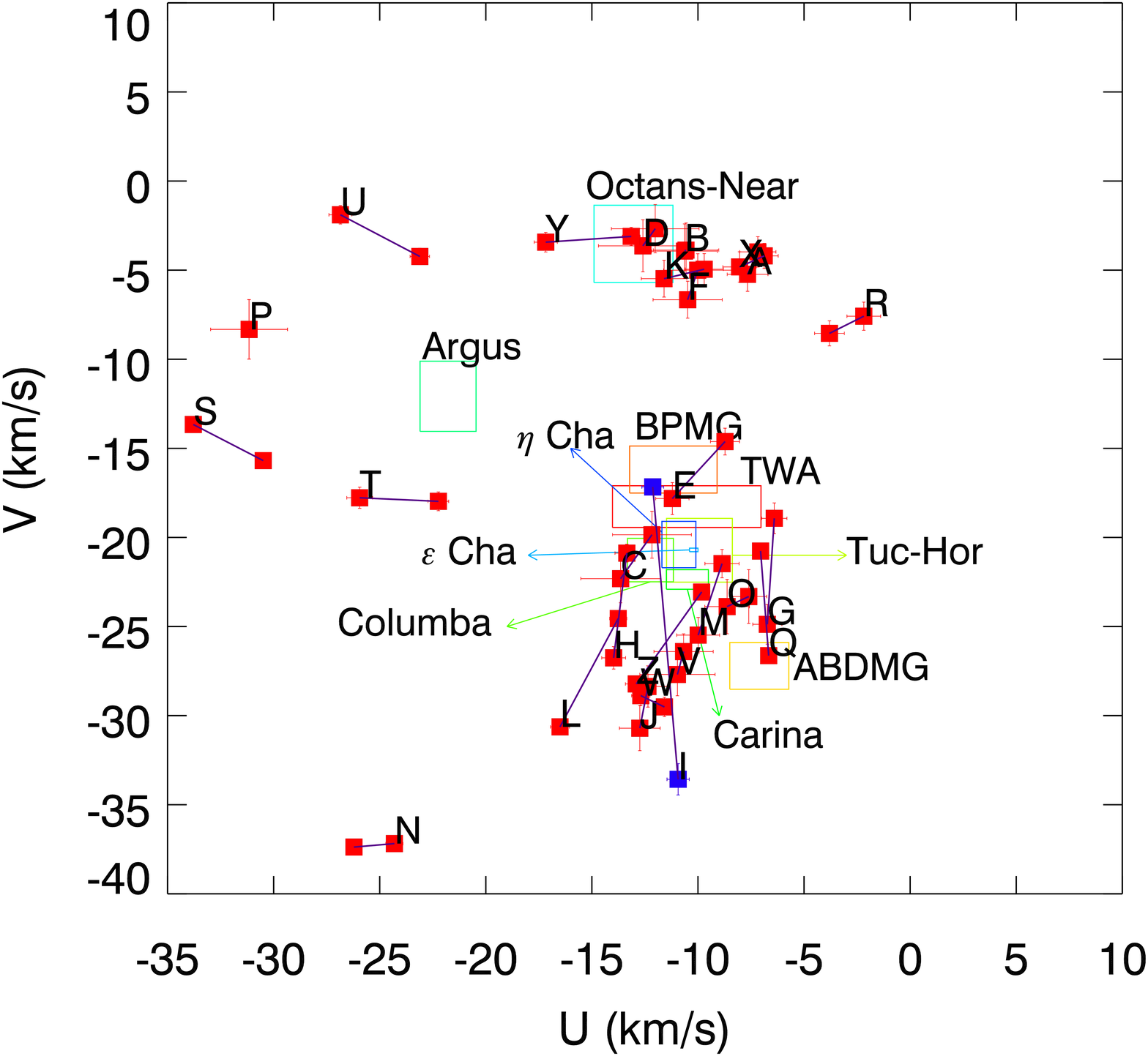}
         \end{minipage}
    \vspace{0.00mm}
        \begin{minipage}{1.0\linewidth}
           \centering
    \vspace{20.00mm}
            \includegraphics[width=1.0\textwidth]{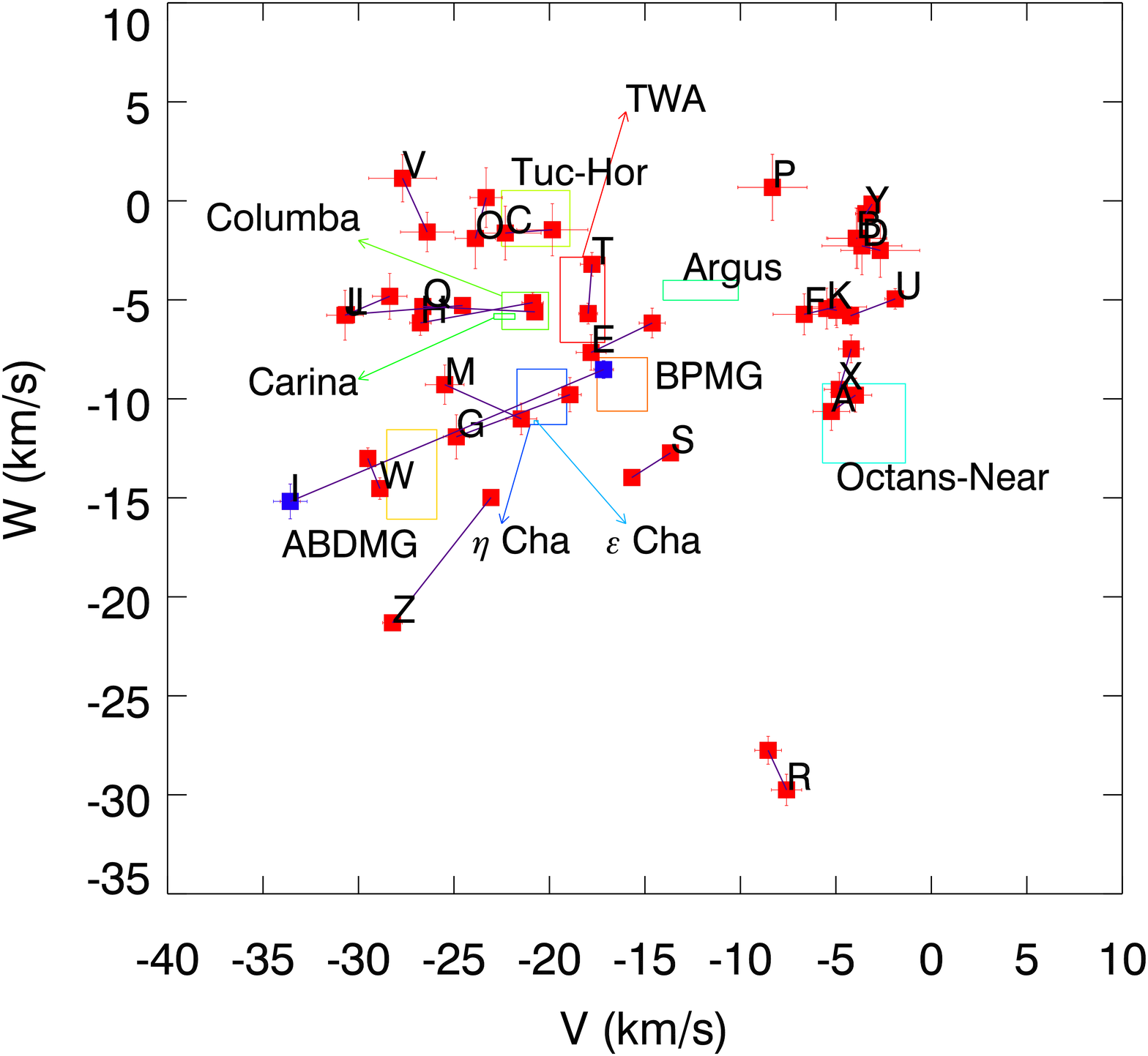}
         \end{minipage}
    \end{center}
        \begin{flushleft}
 \caption{Boettlinger diagrams in $U-V$ and $V-W$ space highlighting the space motions of the characteristically young, likely-single objects in this paper. Each object is connected by a solid purple line which indicates distance uncertainties due to the photometric parallax method described in $\S$\ref{S_Parallax}. Labels for the likely-young sample are the same as in Table~\ref{T_PX_LYS}.}
      \label{F_UVW}
    \end{flushleft}
    \end{figure}

In Figure~\ref{F_UVW} the likely-young sample are plotted on Boettlinger $U-V$ and $V-W$ diagrams. Their space motions are compared with 10 nearby MGs (the $UVW$ used to define one sigma errors for the MGs are taken from table 1 in \citealt{2014a_Gagne} and are listed in Table~\ref{T_UVW_MG}). With the exception of target P, which has a measured parallax, velocity error bars on this plot do not include an error due to the parallax uncertainty. Instead, a line connects $UVW$ points calculated at the extrema of the distances inferred from the photometric parallaxes. The object plotted in blue is target I which had an age range significantly younger than the rest of the sample (see Table~\ref{T_Ages_LYS}). We also note that our analysis does not consider the possibility that our targets could be unresolved binaries which would enlarge our parallax errors.

\input{Kin_LYS}
\input{UVW_MG}

\subsection{Comparison with MG space velocities}\label{S_Chisq}

To test if a star is kinematically matched in $UVW$ to a known MG, a reduced $\chi^{2}$ fitting statistic is used, as described in \cite{2012a_Shkolnik}:

\begin{equation}
 \bar\chi_{\rm T}^{2} = \frac{1}{3} \left(\frac{(U_{*}-U_{\textrm{MG}})^{2}}{\sigma_{U_{*}}^{2} + \sigma_{U_{\textrm{MG}}}^{2}} + \frac{(V_{*}-V_{\textrm{MG}})^{2}}{\sigma_{V_{*}}^{2} + \sigma_{V_{\textrm{MG}}}^{2}} + \frac{(W_{*}-W_{\textrm{MG}})^{2}}{\sigma_{W_{*}}^{2} + \sigma_{W_{\textrm{MG}}}^{2}} \right)
\label{E_Chisq}
\end{equation}

With the exception of target P, the $UVW$ midpoint values are used as inputs for $UVW$ and half the line length is used for the errors. Table~\ref{T_Chisq} provides the reduced $\bar\chi_{\rm T}^{2}$ values for all of the likely-young sample. To interpret these results, given there are 3 degrees of freedom, a $\bar\chi_{\rm T}^{2}$ value larger than 3.78 rejects the null hypothesis at 99 per cent confidence. Any candidate/MG test where $\bar\chi_{\rm T}^{2} > 3.78$ results in rejection for membership of that MG.

\subsection{Predicted radial velocity comparisons with known MGs}\label{S_RVcos}

The Boettlinger diagrams require distances, which, with one exception for the likely-young sample, could only be inferred using photometric parallaxes. Therefore a \textit{distance independent} criterion was used to strengthen the case for MG membership. We have recalculated the convergent points of each MG using the membership lists provided by \cite{2014a_Gagne}, \citealt{2013a_Zuckerman} (for Octans-Near) and \citealt{2013a_Murphy} (for $\epsilon$ and $\eta$~Cha). Given the $UVW$ and convergent point of an MG and the position of a candidate member, the predicted RV for a group member is then $V_{T}\cos\lambda$, where $V_{T}$ is the total velocity of the group ($\sqrt{U^{2} + V^{2} + W^{2}}$) and $\lambda$ is the angle between the convergent point and the target position.

The average RV error for the likely-young sample is $\sim 1\,{\rm km\,s}^{-1}$ and the dispersion amongst MG members in $UVW$ is $\lesssim 3\,{\rm km\,s}^{-1}$. Therefore true MG members are required to have RVs that are within $5\,{\rm km\,s}^{-1}$ of the predicted RV ($\Delta {\rm RV} = |{\rm RV} - V_{T}\cos\lambda| < 5\,{\rm km\,s}^{-1}$). Any candidates with $\Delta {\rm RV} > 5\,{\rm km\,s}^{-1}$ are rejected as MG members. Objects that have RVs within this threshold are flagged as possible members, although even if they satisfy this criterion, a discrepancy in distance, age or chemical abundances may result in a rejection. For this reason the technique can only be used to eliminate membership. The convergent point of each MG (following the prescription in \citealt{1971a_Jones}) is given in Table~\ref{T_UVW_MG} and values of $\Delta {\rm RV}$ for each candidate/MG are given in Table~\ref{T_RV_Proj}.

In $\S$\ref{S_Bin_Sim}, it was calculated that there is a 7.2 per cent chance that a star has 2 RV measurements within $5\,{\rm km\,s}^{-1}$ but is still $> 5\,{\rm km\,s}^{-1}$ from the average centre-of-mass RV. This implies that out of a sample of 26 Li-rich stars likely to be single based on repeat RV measurements, $\sim 2$ may be shifted in $UVW$ by $\sim 5\,{\rm km\,s}^{-1}$ which would potentially alter our assessment of their membership status for any MG.

In principle we could perform a similar test using the tangential velocities, if the distance and proper motion are known accurately (i.e. $\delta V_{\rm TAN} < 5\,{\rm km\,s}^{-1}$). In practice we can only do this on the one object -- target P, which is within $4.6\,{\rm km\,s}^{-1}$ of ABDMG but $> 5\,{\rm km\,s}^{-1}$ for all other MGs.

\subsection{Combined MG membership criteria}\label{S_MG_Criteria}

In this work a candidate only remains a possible member of a MG if it satisfies all of the following criteria:

\begin{enumerate}
 \item The $\bar\chi^{2}_{T}$ value in Table~\ref{T_Chisq} is no more than 3.78.
 \item $\Delta {\rm RV}$ values in Table~\ref{T_RV_Proj} are no more no than $5\,{\rm km\,s}^{-1}$.
 \item The estimated age range for the candidates (provided in Table~\ref{T_Ages_LYS}) must overlap with the age range of the MG, which are provided in a supplementary online table.
\end{enumerate}

In Table~\ref{T_MG_Test} the outcome for each criteria is given. A \tm\ indicates a successful criterion match and an \xm\ represents a failure.

\input{Chisq}
\input{RV_Proj}
\input{MG_Test}

\subsection{Comparison with BANYAN}\label{S_BANYAN}

\cite{2013a_Malo} and \cite{2014a_Gagne} have presented tools for the Bayesian Analysis of Nearby Young AssociatioNs (BANYAN I and BANYAN II respectively). These tools attempt to assign each star a probability of membership for seven well-defined MGs (TWA, BPMG, Tuc-Hor, Columba, Carina, Argus, ABDMG) using sky position, proper motion, photometry and, where available, radial velocities and parallaxes, as inputs. BANYAN II is an improvement over BANYAN I in terms of its treatment of uncertainties and a more realistic assessment of the likely field star contamination using the Besan\c{c}on Galactic model (\citealt{2012a_Robin}).

We have used the provided web-based tools\footnote{available at {\tt http://www.astro.umontreal.ca/\~malo/banyan.}} to test MG membership for our sample of 26 likely young objects. Using the BANYAN I technique, almost all our targets have zero or very low probability of membership for any of the tested MGs. In BANYAN II, the more realistic incorporation of the field population results in even lower probabilities; none of our young targets has a MG membership probability of more than 2 per cent.

The results from BANYAN appear to contradict the assessments we have made in $\S$\ref{S_MG_Criteria}. The reason appears to be a philosophical difference in how MG membership is defined. BANYAN requires that new MG members have similar spatial coordinates to those members that are already known. The predominantly northern hemisphere locations of our targets mean that their $XYZ$ positions are far from the centres of most of the MGs (as defined in BANYAN). Consequently, the membership probabilities are very low. However in this paper we choose to define members of MGs only on the basis of their youth and kinematics and so do not consider the BANYAN membership probabilities any further.

\section{Individual objects}\label{S_Indiv}

\subsection{Individual objects linked to at least one MG}\label{S_Indiv_MG}

Here we focus on objects that satisfied all 3 criteria in $\S$\ref{S_MG_Criteria} for at least one MG. The following section discusses any of these objects that have literature sources pertaining to MG membership and we compare them to our findings in this work.

\noindent\textbf{Target G} - All three criteria are successfully matched with ABDMG. Although target G satisfies both the $\bar\chi_{T}^{2}$ and RV criteria, we have not assigned it to BPMG because its age is measured between 30 and 150\,Myr, too old to be a BPMG member (21--26\,Myr, see Table~\ref{T_UVW_MG}). The age of this object is more likely to be coeval with ABDMG (between 70 and 125\,Myr). There is an uncertainty of $\sim 4\,{\rm km\,s}^{-1}$ in $V$ velocity, which means the object could have kinematic matches to Tuc-Hor or TWA (but not BPMG). A literature search for this object using the SIMBAD\footnote{\url{http://simbad.u-strasbg.fr/simbad/}} database reveals that it was classified as a Pleiades member (Pels 22) by \cite{1986a_van_Leeuwen} and has since been used as a Pleiades member in at least 7 subsequent publications. Target G is very close to the centroid of the Pleiades cluster ($\alpha$, $\delta$ = 03h 47m 00s, +24d 07m 00s) and it has a RV consistent to within $1\,\sigma$, however the parallax range used in this work ($8.78-11.37$\,mas) does not encompass the Pleiades ($7.34 \pm 0.06$\,mas, \citealt{2014a_Melis}), which casts doubt on its membership. If target G were an equal mass binary, an increase of 0.75\,mag (half the flux) may resolve the distance discrepancy.


\noindent\textbf{Target I} - Although this K3 spectral-type object matches all three criteria for seven MGs (BPMG, Tuc-Hor, Columba, Carina, Octans, $\eta$~Cha and $\epsilon$~Cha), there is a large uncertainty in $V$ velocity. Figure~\ref{F_UVW} shows that membership with Tuc-Hor, Carina, Columba and Octans-Near cannot be ruled out, however a more accurate parallax is required to identify which group it may belong to. The object has an Li EW ($> 300$\,m\AA), large enough to set an age $< 30\,$Myr. \cite{1998a_Li} identify this object as a weak-lined T-Tauri star (WTTS) in the surrounding area of the Taurus-Auriga region. It is classified as a T-Tauri star in the pre-main sequence catalog developed by \cite{2005a_Ducourant}, however no excess emission indicative of circumstellar material is observed at near-infrared wavelengths. The distance to Taurus (140\,pc, $\sim 7.1\,$mas, \citealt{2007a_Gudel}) is within our photometric parallax range of $6.46-12.74\,$mas.

\cite{2008a_Santos} located the object in the Taurus SFR and derive chemical abundances. To test for abundance matches with MGs, the mean and standard deviations of [Fe/H], [Si/H] and [Ni/H] for members of the MGs in tables 2 and 3 in \cite{2009a_Viana_Almeida} were compared to target I. The abundances are consistent with any of the considered MGs to within $1\,\sigma$ uncertainties. \cite{2012a_Biazzo} identify this target as part of a $2-10$\,Myr stellar aggregrate within $15^{\circ}$ of Taurus-Auriga, which they name `X-Clump 0534+22'. 


\noindent\textbf{Target Z} - This target is matched with ABDMG and membership is rejected for TWA because target Z is too old to be coeval with TWA. The Li EW/$V-K_{\rm s}$ of the target sets an age range of $30-150$\,Myr (see Figure~\ref{F_Li_EW}). \cite{2000a_Li} identify the object as a possible WTTS and \cite{2005a_Ducourant} include it as a member in the PMS stars proper-motion catalog. The $UVW$ values all lie outside the ABDMG box in Figure~\ref{F_UVW}, however, the $UVW$ uncertainties from parallax estimation are large. A more precise age and parallax are necessary to test if it is an ABDMG member.

\subsection{Individual objects not linked to any MG}\label{S_Indiv_no_MG}

\noindent\textbf{Target B} - At a spectral-type of G1 and Li EW of 154\,m\AA\ this target is assigned an age range of $30-200$\,Myr, supported by a lower age limit of 20\,Myr using H$\alpha$ measurements. In terms of $UVW$ velocity, it appears consistent in $U-V$ with Octans-Near, however the $W$ velocity of the target is $\sim 10\,{\rm km\,s}^{-1}$ larger than Octans-Near. \cite{1995a_Favata} measure a $v\sin i$ value of $34\,{\rm km\,s}^{-1}$ (no error bars provided), which is $1.5\,{\rm km\,s}^{-1} (3\sigma)$ slower than the measurement in this work.

\noindent\textbf{Target D} - The age measured for the target is $30-200$\,Myr based on Li measurements. This age is supported by a lower age limit of 20\,Myr set by H$\alpha$ measurements. It is in the range of Octans-Near members in terms of $U$ and $V$ velocity, but is $\sim 10\,{\rm km\,s}^{-1}$ larger in $W$. It is flagged as a candidate WTTS in Taurus-Auriga region in \cite{1998a_Li} and our photometric parallax range is consistent with this idea.

\noindent\textbf{Target F} - Age estimates are $30-200$\,Myr from Li measurements, and $> 25\,$Myr based on H$\alpha$ measurements. It appears consistent with the $U$ and $V$ velocities of Octans-Near, but is around $5\,{\rm km\,s}^{-1}$ larger in $W$.

\noindent\textbf{Target J} - An age range of $30-150$\,Myr was estimated from Li measurements, consistent with a lower age limit of 20\,Myr set by H$\alpha$. It is $> 5\,{\rm km\,s}^{-1}$ in $U$, $V$ and $W$ from any of the 10 MGs considered.

\noindent\textbf{Target N} - Although target N has an Li EW of 86\,m\AA, because it is a late-K dwarf it is estimated to have an age of $30-150$\,Myr, with an upper limit of 189\,Myr set by H$\alpha$ measurements. The $W$ velocity is matched to a few MGs, however, both the $U$ and $V$ are at least $\sim 15\,{\rm km\,s}^{-1}$ from any MG.

\noindent\textbf{Target P} - Using Li, an age of $30-150$\,Myr was estimated and H$\alpha$ measurements suggest an age of $65-125$\,Myr. It has a $U, V$ and $W$ velocity at least $10\,{\rm km\,s}^{-1}$ from any MG. \cite{2012a_Xing} include this target as a young K4V object (we calculate K5) with a Li EW of 17\,m\AA\, which is almost 150\,m\AA\ lower than the measurement in this work. It is listed as a possible visual binary in \cite{2003a_Makarov}, however, two consistent RV measurements in this analysis suggest the object is not a close binary.

\noindent\textbf{Target Q} - The age range of $30-150$\,Myr may be more consistent with the pattern observed in IC~2602 given the spectral-type K5 and Li EW of 174\,m\AA. The H$\alpha$ age is estimated as 40--65\,Myr. The $U$ and $V$ velocities are consistent with ABDMG, however its $W$ velocity is too large by $\sim 10\,{\rm km\,s}^{-1}$.

\noindent\textbf{Target R} - An age of $30-150$\,Myr constrained by Li measurements is supported by an age of $> 20$\,Myr from H$\alpha$. It is at least $10\,{\rm km\,s}^{-1}$ from any MG in either $U$, $V$ or $W$.

\noindent\textbf{Target S} - Because target S has a late-K spectral-type, its Li EW (135\,m\AA) leads to an age of $30-150$\,Myr, supported by a H$\alpha$ age of $< 230$\,Myr and an $R\sin i$ age of $< 80$\,Myr. It is $> 5\,{\rm km\,s}^{-1}$ from any MG in $U$, $V$ or $W$.

\noindent\textbf{Target T} - Its Li EW is only 62\,m\AA, however, as it is a late-K type, it was estimated to have an age of $100-200$\,Myr. This is too old to be a BPMG member. An limit of $< 120$\,Myr is provided because H$\alpha$ is in emission. There are no matches $< 5\,{\rm km\,s}^{-1}$ in all $UVW$ velocities for any MG.

\noindent\textbf{Target U} - This target has an age range of $30-150$\,Myr, however given the large Li EW (278\,m\AA), this is more likely towards the younger end of this range. It has a H$\alpha$ age older than 35\,Myr. There are no matches within $5\,{\rm km\,s}^{-1}$ for $U, V$ or $W$ for any MG and BPMG membership can be ruled out based on both the kinematics and age.

\noindent\textbf{Target V} - The age range is estimated as $30-150$\,Myr from Li and $< 50\,$Myr based on H$\alpha$ emission. Target V is not within $5\,{\rm km\,s}^{-1}$ in $U$, $V$ or $W$ for any MG.

\noindent\textbf{Target W} - This object satisfies the RV criteria in $\S$\ref{S_RVcos} for ABDMG (see Table~\ref{T_RV_Proj}) and its location in $U-V$ and $V-W$ velocity space suggests that it is similar to ABDMG in $V$ and $W$ but is more than $5\,{\rm km\,s}^{-1}$ from ABDMG in $U$ velocity and therefore fails membership test 1, although only marginally ($\bar\chi_{T}^{2} = 4.5$, the membership criteria is $\bar\chi_{T}^{2} < 3.78$). The $U, V$ and $W$ are precise to within $1\,{\rm km\,s}^{-1}$ in each velocity coordinate. The Li EW measurement of 222\,m\AA\ at an early-K spectral-type results in an age range of 30 to 150\,Myr and although there is significant scatter amongst IC~2602 and Pleiades objects at this spectral-type, this target is more consistent with the lower envelope of IC~2602. The H$\alpha$ age range is $35-55$\,Myr and an age of $< 80$\,Myr is implied by $R\sin i$.

\cite{2011a_Frasca} identify differential rotation based on data from the \textit{Kepler} mission. Their Li EW measurement is $\sim 50$\,m\AA\ larger and their 1.0\AA\ H$\alpha$ EW is certainly not consistent with the 0.1\AA\ absorption line measured in this work. The $v\sin i$ measurements match to 1$\sigma$ and their calculated inclination angle, $i$ ($\simeq 70^{\circ}$), provides a stellar radius of 0.93$R_{\odot}$, which from Figure~\ref{F_R_VK} suggests an age of$ 25-80$\,Myr, where the $\sin i$ ambiguity has been resolved. This is at least 40\,Myr younger than ABDMG (if indeed ABDMG is coeval with the Pleiades), making membership for ABDMG unlikely. The authors also note that the object is flagged as an equal-magnitude visual binary system in the WDS Catalog (WDS~19251+4431, \citealt{1997a_Worley}) and the Tycho Double Star Catalog \citep{2002a_Fabricius}. Based on the orbital properties calculated by \cite{2011a_Frasca}, if it were an equal-mass binary it would be expected to have RV variations $> 5\,{\rm km\,s}^{-1}$ over $3-4$ years. The RVs reported by \cite{2011a_Frasca} were from 3 separate observations between 2007 and 2009 ($-33.1, -33.1, -32.2\,{\rm km\,s}^{-1}$) and the measurement of $-33.5 \pm 0.6\,{\rm km\,s}^{-1}$ in this work carried out in June 2011 suggests that the object is not in a equal-mass visual binary system.

\noindent\textbf{Target Y} - This target has a spectral-type K4 and a Li EW of 141\,m\AA, from which an age range of $30-150$\,Myr was estimated, and a H$\alpha$ emission line implies an age of $< 100$\,Myr. It is within the range of Octans-Near in $U$ and $V$ velocity, but has a $W$ velocity $\sim 5\,{\rm km\,s}^{-1}$ larger than Octans-Near.

\subsection{Connection with the Octans-Near association?}\label{S_Octans}

In a Hipparcos-based survey, \cite{2013a_Zuckerman} report 14 star systems with spectral-types between G5 and A0 and distances ranging from 24 to 98\,pc that have ages between $\sim 30-100$\,Myr and Galactic space velocities similar to the mean space motion of Octans ($UVW_{\rm Octans} = -14.5 \pm 0.9, -3.6 \pm 1.6, -11.2 \pm 1.4\,{\rm km\,s}^{-1}; UVW_{\rm sample} = -13.0 \pm 1.9, -3.5 \pm 2.2, -11.2 \pm 2.0\,{\rm km\,s}^{-1}$). \cite{2008a_Torres} measure the distance of Octans to be 141 $\pm$ 34\,pc and given that the 14 star systems in \cite{2013a_Zuckerman} are closer than the Octans association (all within 100\,pc), the authors describe this stellar aggregate as the 'Octans-Near' group. 

In $\S$\ref{S_MG_Criteria}, four objects -- A, I, K and X -- pass tests for Octans-Near membership, however, target I is unlikely to be an Octans-Near member because the $UVW$ error bars are so large that the fact it agrees with Octans-Near is not remarkable (see $\S$\ref{S_Indiv_MG}). Whether these objects are connected to either the Octans, the Octans-Near or not, there nevertheless still exists a sub-grouping of 7 objects in the sample (A, B, D, F, K, X and Y) which are possibly coeval. All 7 objects could be consistent with a common age of between 30--150\,Myr, which is also consistent with Octans-Near. These have $UVW$ values of $U = -10.5 \pm 2.7 (\pm 1.8); V = -4.3 \pm 1.0 (\pm 1.4); W = -4.9 \pm 3.6 (\pm 1.3)\,{\rm km\,s}^{-1}$, which is close to Octans/Octans-Near but not to any other known MG (error bars are the standard deviations and the values in parentheses are the average $UVW$ uncertainties). Of these seven, none are in similar in terms of $XYZ$ positions, although five of them, A, B, D, F and Y, have sky positions within 3 hours of right ascension of one another and 10 degrees in declination. These are in the vicinity of the Pisces constellation, therefore if these did constitute a MG and were not associated with Octans-Near, we propose naming the ensemble as ``the Pisces moving group''.

In principle, if the 7 kinematically consistent objects identified here were born from the same molecular cloud, then given their $UVWXYZ$ coordinates, it would be possible to trace-back their motions under a reasonable gravitational potential and identify a time at which all these objects occupied a minimum. However, since our objects all have designated ages older than 30\,Myr and lack trigonometric parallaxes, a kinematic traceback analysis is premature: it would only take 5\,Myr for the current velocity uncertainties to produce a spatial dispersion equal to the current standard deviation of stellar positions.

\section{Conclusions}\label{S_Conclusions}

From an initial sample of 146 spectroscopically observed, FGK stars with short rotation periods from the SuperWASP All Sky Survey, 26 were found to be Li-rich and therefore probably young by comparison with the photospheric Li abundance patterns seen in open clusters. Duplicate RV measurements suggest these are unlikely to be close binaries. Five objects were G-stars and the remaining 21 have spectral types from K0 to K6.

Twelve objects from the initial observed target sample had 2 or more RV measurements that varied by $> 5\,{\rm km\,s}^{-1}$ on the timescale of the observing run and are almost certain to be close, tidally-locked binary systems. This fraction of close binary systems is much higher than expected from a random sample of field stars, presumably because of our initial selection based on rapid rotation. If a significant Li line was not observed at the telescope, then usually a repeat measurement was not made. Of the remaining Li-depleted objects, for which only a single RV measurement was obtained, we anticipate that a large fraction will turn out to be close binary systems.

Typical ages of members in the likely-young sample are 30--200\,Myr and have distances of 30--180\,pc. Ages are primarily determined from lithium, but ages derived from rotation, activity and projected radius are usually in agreement, but might be compromised by binarity. Lower-mass, nearby, young stars are favourable as targets to follow-up with high-resolution imaging to detect the presence of brown-dwarf binaries, or even exoplanets. There are 16 objects in the likely-young sample with ages $< 200$\,Myr, later than K0 and within 100\,pc that would be ideal candidates for imaging surveys.

Fifty-four per cent of the observed sample had rotation periods $< 2$ days (77 out of 146) and there were fourteen objects in the likely-young sample that had rotation periods less than 2 days. The sample may not be complete since sometimes even young stars do not exhibit significant rotational modulation. In terms of the distribution of rotation periods, the likely-young sample and parent sample have similar medians, therefore one would not gain by being more restrictive in choosing a rotation period cut-off.

Eleven targets are identified which satisfy kinematic and age criteria with at least one MG and fifteen fail at least one criteria for membership of \textit{any} of the MGs considered in this analysis. There is a low probability of the RVs being significantly in error due to binary motion (see $\S$\ref{S_Bin_Sim}), therefore it is unlikely that many of these classifications would change with more extensive RV observations. There is tentative evidence for a sub-grouping of seven objects that appear to be a northern hemisphere counterpart to, but not identical to, the `Octans-Near' MG proposed by \cite{2013a_Zuckerman}. Regardless of whether some of these objects are connected to Octans-Near or not, a grouping of stars exist around $U = -10.5 \pm 2.7; V = -4.3 \pm 1.0; W = -4.9 \pm 3.6\,{\rm km\,s}^{-1}$ which we tentatively label as ``the Pisces moving group''.

In light of results from the analysis of these 146 objects, additional substantial surveys of MG candidates in the northern hemisphere are desirable. The kinematically unbiased search mechanism was $\sim 18$ per cent efficient at detecting likely-single G and K-type stars younger than 200\,Myr and can be used to identify stars that are not associated with previously studied MGs. More widespread searches for MG candidates in the northern hemisphere may provide additional useful samples of young stars, and could unveil important kinematic substructure in the Solar neighbourhood.

\section{Acknowledgements}

We would like to thank the anonymous referee for a prompt review and for the comments which have greatly helped to improve the manuscript. Thanks are due to the staff at the NOT for their help during the observation. Particular thanks are owed to Eric Stempels for his use of the FIEStool package and S{\o}ren Frimann who gave useful advice on using the data reduction pipeline. Thanks are also due to John Webb for constructing a comprehensive initial data search on the target stars, which served as a very useful reference. This research has made use of the SIMBAD database, operated at CDS, Strasbourg, France. This publication makes use of data products from the Two Micron All Sky Survey, which is a joint project of the University of Massachusetts and the Infrared Processing and Analysis Center/California Institute of Technology, funded by the National Aeronautics and Space Administration and the National Science Foundation.

\bibliography{Paper}
\label{lastpage}
\end{document}

%% file: P_Xray.tex
{\scriptsize
\begin{table*}
  \caption{Measured periods for the sample of 26 likely-young objects (see $\S$\ref{S_Age_Final}).  Periods are the reanalysed values (and standard errors) using the analysis in $\S$\ref{S_Period}.  The first error bar is the (averaged) uncertainty calculated using equation 2 in \protect\cite{2010a_Messina}.  The second error bar (where appropriate) is the standard error in measurements for 2 or more seasons. `$N$' refers to the number of seasonal lightcurves analysed for each target star.  The column labelled `$Q$' refers to the quality of the period determination. The `Date' column refers to the month and year when the first observation of the target in this work took place, J11 = June 2011, and D12 = December 2012. X-ray count rates (CR) and HR1 ratios are extracted from either the 1RXS or 2RXP catalogs and $\log L_{X}/L_{\rm bol}$ is the X-ray to bolometric luminosity ratio. The equivalent data for all other objects observed in this work (and for all subsequent tabular data) are available online.}
\begin{center}
\begin{tabular}{lrrlrrrrrrrrr}
\hline
\hline
SuperWASP ID/Label            & $V$    & $V-K_{\rm s}$ &  Period         & $\Delta\chi^{2}$ & $N$ & $Q$ & Date & CR & HR1 & $\log L_{\rm X}/L_{\rm bol}$ \\
(1SWASP J-)                   & (mag)  & (mag) &  (days)         &                  &     &     &      &    &     &                              \\
\hline
012457.96+255702.4, A         & 10.718 & 1.811 & 3.048$\pm$0.120$\pm$0.026 & 1711  & 2  & B & D12 	& 0.06  &    0.42 & $-$3.35 \\
013514.32+211622.4, B         & 10.788 & 1.490 & 1.871$\pm$0.010$\pm$0.004 & 1994  & 3  & A & D12 	& 0.01  &    1.00 & $-$3.84 \\
023503.81+313922.1, C         & 10.337 & 1.743 & 1.274$\pm$0.013$\pm$0.002 & 2038  & 4  & A & D12 	& 0.15  &    0.17 & $-$3.18 \\
030405.14+300309.6, D         & 10.875 & 1.807 & 1.806$\pm$0.020$\pm$0.005 & 1518  & 2  & B & D12 	& 0.06  &    0.66 & $-$3.23 \\
031628.14+563857.7, E         & 10.667 & 2.103 & 3.850$\pm$0.080           & 584   & 2  & C & D12 	& 0.08  &    0.34 & $-$3.31 \\
032231.55+285319.8, F         & 10.672 & 2.013 & 1.654$\pm$0.016$\pm$0.011 & 281   & 2  & C & D12 	& 0.06  &    0.86 & $-$3.34 \\
034319.02+222657.2, G         & 11.571 & 2.545 & 2.893$\pm$0.036           & 1075  & 1  & B & D12 	& 0.06  &    1.00 & $-$3.07 \\
050206.19+311102.2, H         & 10.792 & 2.218 & 4.531$\pm$0.116           & 4169  & 1  & B & D12     & 0.07  & $-$0.07 & $-$3.48 \\
052146.83+240044.4, I         & 10.393 & 2.323 & 3.436$\pm$0.054           & 9955  & 1  & C & D12     & 0.30  & $-$0.08 & $-$3.01 \\
084748.63+342356.8, J         & 10.304 & 1.686 & 4.231$\pm$0.074           & 996   & 2  & C & D12 	& 0.03  & $-$0.08 & $-$3.90 \\
133708.20+444454.5, K         & 11.050 & 1.954 & 4.376$\pm$0.196           & 301   & 1  & C & D12 	& 0.07  & $-$0.11 & $-$3.32 \\
135458.37$-$054354.1, L       & 9.231  & 2.091 & 5.361$\pm$0.140$\pm$0.040 & 319   & 2  & C & D12 	& 0.09  & $-$0.60 & $-$4.16 \\
143854.57+330019.9, M         & 11.036 & 2.219 & 5.095$\pm$0.156$\pm$0.036 & 3576  & 2  & B & D12 	& 0.05  &    0.40 & $-$3.36 \\
162506.55+300225.8, N         & 10.205 & 3.022 & 1.002$\pm$0.006$\pm$0.098 & 51786 & 2  & B & D12 	& 0.22  &    0.29 & $-$3.27 \\
214537.36+271110.8, O         & 11.425 & 2.305 & 1.267$\pm$0.015$\pm$0.225 & 1152  & 3  & B & D12 	& 0.06  &    0.29 & $-$3.23 \\
133241.69+223006.6, P         & 9.655  & 2.808 & 1.040$\pm$0.013           & 186   & 1  & C & J11 	& 0.84  & $-$0.13 & $-$2.97 \\
155007.38$-$022211.5, Q       & 10.050 & 2.538 & 2.241$\pm$0.019$\pm$0.007 & 37160 & 3  & A & J11 	& 0.48  & $-$0.09 & $-$2.99 \\
162946.59+281038.0, R         & 10.654 & 1.925 & 1.426$\pm$0.011$\pm$0.003 & 5023  & 4  & A & J11 	& 0.09  &    0.87 & $-$3.14 \\
171808.56+250612.0, S         & 10.614 & 3.027 & 2.420$\pm$0.047$\pm$0.008 & 5432  & 2  & B & J11 	& 0.27  &    0.04 & $-$3.08 \\
172228.64+365842.1, T         & 10.504 & 2.800 & 1.229$\pm$0.011$\pm$0.001 & 10864 & 9  & A & J11 	& 0.25  & $-$0.06 & $-$3.14 \\
173103.32+281506.5, U         & 10.213 & 2.402 & 1.263$\pm$0.012$\pm$0.002 & 13828 & 3  & A & J11 	& 0.29  & $-$0.05 & $-$3.11 \\
180426.56+393047.1, V         & 11.532 & 2.365 & 1.547$\pm$0.017$\pm$0.004 & 3260  & 5  & A & J11 	& 0.07  &    0.01 & $-$3.15 \\
192502.00+442950.7, W         & 9.939  & 2.398 & 1.205$\pm$0.013$\pm$0.001 & 11141 & 2  & B & J11 	& 0.24  &    0.18 & $-$3.23 \\
205830.74$-$090223.3, X       & 11.180 & 2.071 & 2.251$\pm$0.026$\pm$0.012 & 5173  & 6  & A & J11 	& 0.11  &    0.28 & $-$2.99 \\
225617.59+205236.2, Y         & 11.410 & 2.731 & 1.100$\pm$0.009$\pm$0.003 & 1164  & 4  & A & J11 	& 0.08  &    0.06 & $-$3.21 \\
230752.70+171015.2, Z         & 10.780 & 2.881 & 1.073$\pm$0.007$\pm$0.002 & 980   & 3  & C & J11 	& 0.30  & $-$0.25 & $-$3.01 \\
\hline
\end{tabular}
\end{center}
\label{T_PX_LYS}
\end{table*}
}

%% file: RV_Templates.tex
{\tiny
\begin{table}
\caption{RV and minimum activity standards - All RV are from \protect\cite{2012a_Chubak}, except for HD~114762 (\protect\citealt{1999a_Udry}).  All $v\sin i$ values are from \protect\cite{2005a_Glebocki}.  All $\log R^{'}_{\rm HK}$ are from \protect\cite{2010a_Isaacson}, except for HD~3651 (\protect\citealt{2013a_Pace}) and HD~190007 (\protect\citealt{2011a_Mittag}).}
\begin{center}
\begin{tabular}{L{1.3cm}R{1cm}R{0.9cm}R{0.9cm}R{0.9cm}R{0.9cm}}
\hline
\hline
Standard      & SpT           & $RV$               & $\sigma_{RV}$      & $v\sin i$          & $\log R^{'}_{\rm HK}$ \\
              &               & ${\rm km\,s}^{-1}$ & ${\rm km\,s}^{-1}$ & ${\rm km\,s}^{-1}$ &                       \\
\hline
HD~114762     & F8V           &    49.4            & 0.5                & 1.7                & $-$4.902              \\
HD~1461       & G0V           & $-$10.158          & 0.116              & 5.0                & $-$5.008              \\
HD~95128      & G1V           &    11.293          & 0.108              & 3.1                & $-$4.973              \\
HD~217014     & G2.5IVa       & $-$33.118          & 0.128              & 2.8                & $-$5.054              \\
HD~197076     & G5V           & $-$35.402          & 0.030              & 2.9                & $-$4.872              \\
HD~9407       & G6V           & $-$33.313          & 0.124              & 0.0                & $-$4.986              \\
HD~115617     & G7V           &  $-$7.844          & 0.128              & 0.5                & $-$4.962              \\
HD~101501     & G8V           &  $-$5.464          & 0.105              & 2.3                & $-$4.483              \\
HD~3651       & K0V           & $-$32.940          & 0.042              & 0.6                & $-$4.849              \\
HD~10780      & K0V           &     2.814          & 0.086              & 0.9                & $-$4.700              \\
HD~131509     & K0V           & $-$44.749          & 0.167              & 6.4                & $-$5.142              \\
HD~4628       & K2.5V         & $-$10.229          & 0.030              & 1.6                & $-$4.979              \\
HD~190007     & K4Vk          & $-$30.270          & 0.105              & 2.8                & $-$4.592              \\
HD~209290     & M0V           &    18.275          & 0.120              & 3.8                &                       \\
HD~119850     & M1.5V         &    15.778          & 0.061              & 1.8                &                       \\
GJ~411        & M2.0V         & $-$84.683          & 0.030              & 1.6                &                       \\
GJ~526        & M4.0V         &    15.778          & 0.061              & 1.8                &                       \\
\hline
\end{tabular}
\end{center}
\label{T_RV_Templates}
\end{table}
}

%% file: RVs_LYS.tex
{\tiny
\begin{table}
\caption{Radial velocity measurements. Individual RVs are provided in column 3 and the final averaged RV in column 4.  The INT RV measurements are subscripted with an `i' and are not factored into the final averaged RV for each target.  The column titled `$B$' is the binary score based on the criteria described in $\S$\ref{S_Bin_Con}.}
\begin{center}
\begin{tabular}{lrrrr}
\hline
\hline
Target        & HJD        & $RV_{\rm indiv}$            & $RV_{\rm final}$     & $B$          \\
              & (2450000+) & (${\rm km\,s}^{-1}$)        & (${\rm km\,s}^{-1}$) &              \\
\hline
A             &   6289.422 & 7.2 $\pm$ 0.7               & 7.3 $\pm$ 0.7        & 1            \\
              &   6290.409 & 7.4 $\pm$ 0.7               &                      &              \\
B             &   6289.358 & 5.0 $\pm$ 1.9               & 5.0 $\pm$ 2.0        & 1            \\
              &   6290.427 & 5.0 $\pm$ 2.2               &                      &              \\
C             &   6289.370 & 0.3 $\pm$ 2.5               & 0.3 $\pm$ 2.4        & 1            \\
              &   6290.442 & 0.2 $\pm$ 2.3               &                      &              \\
D             &   6289.476 & 9.8 $\pm$ 2.4               & 9.8 $\pm$ 2.4        & 1            \\
              &   6290.448 & 9.8 $\pm$ 2.4               &                      &              \\
E             &   6289.488 & $-$2.0 $\pm$ 0.6            & $-$2.1 $\pm$ 0.6     & 1            \\
              &   6290.460 & $-$2.1 $\pm$ 0.6            &                      &              \\
F             &   6289.498 & 8.2 $\pm$ 1.8               & 9.0 $\pm$ 1.8        & 1            \\
              &   6290.469 & 9.9 $\pm$ 1.9               &                      &              \\
G$^{\rm a}$   &   6291.569 & 6.0 $\pm$ 0.5               & 6.0 $\pm$ 0.5        & 1            \\
H             &   6291.589 & 11.2 $\pm$ 0.6              & 11.2 $\pm$ 0.6       & 1            \\
              &   6376.436 & 14.2 $\pm$ 2.2$_{\rm i}$    &                      &              \\
I$^{\rm b}$   &   6291.598 & 13.5 $\pm$ 0.5              & 13.5 $\pm$ 0.5       & 1            \\
              &   6376.429 & 17.1 $\pm$ 2.2$_{\rm i}$    &                      &              \\
J             &   6289.580 & 10.1 $\pm$ 0.6              & 10.2 $\pm$ 0.6       & 1            \\
              &   6290.617 & 10.2 $\pm$ 0.6              &                      &              \\
              &   6375.474 & 11.6 $\pm$ 1.5$_{\rm i}$    &                      &              \\
K             &   6289.720 & $-$6.4 $\pm$ 0.6            & $-$6.4 $\pm$ 0.6     & 1            \\
              &   6290.708 & $-$6.4 $\pm$ 0.6            &                      &              \\
              &   6375.619 & $-$1.2 $\pm$ 3.5$_{\rm i}$  &                      &              \\
L             &   6289.731 & $-$4.1 $\pm$ 0.6            & $-$4.2 $\pm$ 0.6     & 1            \\
              &   6290.719 & $-$4.2 $\pm$ 0.6            &                      &              \\
              &   6375.631 & $-$2.3 $\pm$ 1.6$_{\rm i}$  &                      &              \\
M             &   6289.738 & $-$19.2 $\pm$ 0.6           & $-$19.2 $\pm$ 0.6    & 1            \\
              &   6291.634 & $-$19.3 $\pm$ 0.6           &                      &              \\
N             &   6289.780 & $-$68.0 $\pm$ 0.7           & $-$68.5 $\pm$ 0.7    & 1            \\
              &   6290.791 & $-$69.1 $\pm$ 0.7           &                      &              \\
O             &   6290.317 & $-$22.5 $\pm$ 1.8           & $-$22.9 $\pm$ 1.6    & 1            \\
              &   6291.314 & $-$23.4 $\pm$ 1.4           &                      &              \\
P             &   5735.502 & $-$5.1 $\pm$ 0.7            & $-$5.1 $\pm$ 0.7     & 1            \\
              &   6375.594 & $-$5.0 $\pm$ 2.2$_{\rm i}$  &                      &              \\
Q$^{\rm c}$   &   5736.457 & $-$10.3 $\pm$ 0.3           & $-$10.6 $\pm$ 0.4    & 1            \\
              &   5737.503 & $-$10.9 $\pm$ 0.4           &                      &              \\
              &   6375.688 & $-$10.4 $\pm$ 2.3$_{\rm i}$ &                      &              \\
R             &   5735.567 & $-$25.3 $\pm$ 0.8           & $-$25.2 $\pm$ 0.8    & 1            \\
              &   5736.515 & $-$25.1 $\pm$ 0.7           &                      &              \\
              &   6375.699 & $-$20.3 $\pm$ 2.2$_{\rm i}$ &                      &              \\
S             &   5736.483 & $-$34.7 $\pm$ 0.3           & $-$34.8 $\pm$ 0.3    & 1            \\
              &   5737.513 & $-$34.9 $\pm$ 0.3           &                      &              \\
T             &   5735.613 & $-$24.6 $\pm$ 0.6           & $-$25.3 $\pm$ 0.6    & 1            \\
              &   5736.524 & $-$26.0 $\pm$ 0.6           &                      &              \\
U             &   5735.643 & $-$18.2 $\pm$ 0.5           & $-$18.2 $\pm$ 0.5    & 2            \\
V             &   5736.591 & $-$25.5 $\pm$ 0.6           & $-$26.4 $\pm$ 0.6    & 1            \\
              &   5737.569 & $-$27.3 $\pm$ 0.7           &                      &              \\
W$^{\rm d}$   &   5734.615 & $-$33.5 $\pm$ 0.6           & $-$33.5 $\pm$ 0.6    & 1            \\
X             &   5736.652 & $-$3.0 $\pm$ 0.4            & $-$2.7 $\pm$ 0.4     & 1            \\
              &   5737.617 & $-$2.4 $\pm$ 0.4            &                      &              \\
Y             &   5737.676 & $-$2.5 $\pm$ 0.6            & $-$2.5 $\pm$ 0.6     & 2            \\
Z             &   5737.662 & $-$8.4 $\pm$ 0.3            & $-$8.4 $\pm$ 0.3     & 2            \\
\hline
\end{tabular}
\end{center}
a: RV = $6.06 \pm 0.29\,{\rm km\,s}^{-1}$ \citep{2009a_Mermilliod}, b: RV = $14.2 \pm 1.4\,{\rm km\,s}^{-1}$ \citep{2012a_Biazzo}, c: RV = $-13.6 \pm 1.8\,{\rm km\,s}^{-1}$ \citep{2013a_Kordopatis}, d: RV = $-33.1, -33.1, -32.2\,{\rm km\,s}^{-1}$ \citep{2011a_Frasca}.
\label{T_RVs_LYS}
\end{table}
}

%% file: Temp_EW_Abun_LYS.tex
{\tiny
\begin{table*}
  \caption{EWs, $BVK$ photometry, temperatures, spectral-types (SpT) and Li abundances.  Li EW$_{\rm c}$ is the final EW after correcting for the blended Fe~{\sc i} line.}
\begin{center}
\begin{tabular}{lrrrrrrrrr}
  \hline
  \hline
Target & Li EW  & Li EW$_{\rm c}$ & H$\alpha$ EW    & $B$                 & $V$                 & $K_{\rm s}$                 & $T_{\rm eff}$ & SpT  & A(Li)                    \\
       & (m\AA) & (m\AA)          & (\AA)           & (mag)               & (mag)               & (mag)               & (K)           &      &                          \\
  \hline
A      & 223    &  213 $\pm$ 13      &    1.28     & 11.470 $\pm$ 0.045  & 10.718 $\pm$ 0.042  &  8.907 $\pm$ 0.019  & 5430          & G9   &  $ 2.89^{+0.09}_{-0.13}$ \\
B      & 154    &  144 $\pm$ 21      &    2.22     & 11.387 $\pm$ 0.008  & 10.788 $\pm$ 0.020  &  9.298 $\pm$ 0.016  & 5850          & G1   &  $ 2.91^{+0.10}_{-0.14}$ \\
C      & 209    &  197 $\pm$ 20      &    0.99     & 10.980 $\pm$ 0.061  & 10.337 $\pm$ 0.049  &  8.594 $\pm$ 0.017  & 5510          & G8   &  $ 2.88^{+0.11}_{-0.15}$ \\
D      & 193    &  186 $\pm$ 21      &    1.77     & 11.579 $\pm$ 0.087  & 10.875 $\pm$ 0.076  &  9.068 $\pm$ 0.018  & 5440          & G8   &  $ 2.77^{+0.14}_{-0.18}$ \\
E      & 222    &  211 $\pm$ 13      &    0.82     & 11.482 $\pm$ 0.049  & 10.667 $\pm$ 0.023  &  8.564 $\pm$ 0.020  & 5120          & K1   &  $ 2.59^{+0.08}_{-0.12}$ \\
F      & 184    &  174 $\pm$ 25      &    0.71     & 11.356 $\pm$ 0.050  & 10.672 $\pm$ 0.059  &  8.659 $\pm$ 0.019  & 5210          & K1   &  $ 2.52^{+0.13}_{-0.18}$ \\
G      & 267    &  246 $\pm$ 25      &    0.49     & 12.537 $\pm$ 0.056  & 11.571              &  9.026 $\pm$ 0.020  & 4710          & K4   &  $ 2.35^{+0.10}_{-0.18}$ \\
H      & 249    &  238 $\pm$ 16      &    0.80     & 11.671 $\pm$ 0.078  & 10.792 $\pm$ 0.017  &  8.574 $\pm$ 0.021  & 5000          & K2   &  $ 2.61^{+0.08}_{-0.13}$ \\
I$^{\rm a}$ & 386    &  366 $\pm$ 14      & $-$0.05     & 11.367 $\pm$ 0.045  & 10.393 $\pm$ 0.043  &  8.070 $\pm$ 0.021  & 4900          & K3   &  $ 3.18^{+0.09}_{-0.14}$ \\
J      & 165    &  155 $\pm$ 15      &    1.83     & 10.958 $\pm$ 0.023  & 10.304 $\pm$ 0.038  &  8.618 $\pm$ 0.022  & 5590          & G6   &  $ 2.76^{+0.10}_{-0.14}$ \\
K      & 260    &  245 $\pm$ 10      &    1.07     & 11.852 $\pm$ 0.027  & 11.050 $\pm$ 0.022  &  9.096 $\pm$ 0.018  & 5270          & K0   &  $ 2.89^{+0.07}_{-0.11}$ \\
L      & 169    &  159 $\pm$ 12      &    1.98     & 10.089 $\pm$ 0.144  &  9.231 $\pm$ 0.104  &  7.140 $\pm$ 0.020  & 5130          & K1   &  $ 2.38^{+0.15}_{-0.19}$ \\
M      & 106    &   93 $\pm$ 8       &    0.96     & 11.934 $\pm$ 0.055  & 11.036 $\pm$ 0.048  &  8.817 $\pm$ 0.020  & 5000          & K2   &  $ 1.92^{+0.11}_{-0.14}$ \\
N      & 86     &   69 $\pm$ 15      & $-$0.40     & 11.311 $\pm$ 0.032  & 10.132 $\pm$ 0.037  &  7.183 $\pm$ 0.027  & 4710          & K5   &  $ 1.95^{+0.14}_{-0.19}$ \\
O      & 254    &  242 $\pm$ 24      & $-$0.08     & 12.014 $\pm$ 0.032  & 11.279 $\pm$ 0.037  &  9.120 $\pm$ 0.018  & 5060          & K2   &  $ 2.68^{+0.11}_{-0.17}$ \\
P$^{\rm b}$ & 159    &  142 $\pm$ 14      & $-$0.62     & 10.669 $\pm$ 0.043  &  9.655 $\pm$ 0.024  &  6.847 $\pm$ 0.023  & 4390          & K5   &  $ 1.07^{+0.10}_{-0.15}$ \\
Q      & 174    &  155 $\pm$ 23      & $-$0.01     & 10.939 $\pm$ 0.041  & 10.050 $\pm$ 0.058  &  7.512 $\pm$ 0.023  & 4490          & K5   &  $ 1.65^{+0.13}_{-0.18}$ \\
R      & 208    &  196 $\pm$ 18      &    1.14     & 11.394 $\pm$ 0.062  & 10.654 $\pm$ 0.041  &  8.729 $\pm$ 0.020  & 5300          & K0   &  $ 2.70^{+0.11}_{-0.15}$ \\
S      & 135    &  117 $\pm$ 4       & $-$1.22     & 11.533 $\pm$ 0.184  & 10.614 $\pm$ 0.069  &  7.587 $\pm$ 0.020  & 4330          & K6   &  $ 1.31^{+0.12}_{-0.12}$ \\
T      & 62     &   44 $\pm$ 7       & $-$0.34     & 11.601 $\pm$ 0.046  & 10.504 $\pm$ 0.039  &  7.704 $\pm$ 0.017  & 4500          & K5   &  $ 1.01^{+0.12}_{-0.16}$ \\
U      & 278    &  265 $\pm$ 36      &    0.19     & 11.069 $\pm$ 0.026  & 10.213 $\pm$ 0.011  &  7.811 $\pm$ 0.016  & 4830          & K3   &  $ 2.58^{+0.15}_{-0.25}$ \\
V      & 269    &  252 $\pm$ 36      & $-$0.18     & 12.603 $\pm$ 0.066  & 11.532 $\pm$ 0.046  &  9.167 $\pm$ 0.017  & 4860          & K3   &  $ 2.63^{+0.17}_{-0.26}$ \\
W$^{\rm c}$      & 222    &  208 $\pm$ 20      &    0.01     & 11.011 $\pm$ 0.047  &  9.939 $\pm$ 0.061  &  7.541 $\pm$ 0.018  & 4840          & K3   &  $ 2.31^{+0.12}_{-0.17}$ \\
X      & 337    &  279 $\pm$ 42      &    0.76     & 12.208 $\pm$ 0.252  & 11.180 $\pm$ 0.112  &  9.109 $\pm$ 0.021  & 5150          & K1   &  $ 2.93^{+0.23}_{-0.29}$ \\
Y      & 141    &  122 $\pm$ 8       & $-$0.36     & 12.390 $\pm$ 0.071  & 11.410 $\pm$ 0.072  &  8.679 $\pm$ 0.016  & 4560          & K4   &  $ 1.62^{+0.13}_{-0.17}$ \\
Z      & 163    &  146 $\pm$ 30      & $-$1.50     & 11.867 $\pm$ 0.054  & 10.780 $\pm$ 0.024  &  7.899 $\pm$ 0.027  & 4440          & K5   &  $ 1.60^{+0.14}_{-0.21}$ \\
\hline
\end{tabular}
\end{center}
\begin{flushleft}
a: H$\alpha$ EW = $-$2.14\,\AA; Li EW = 390\,m\AA \citep{1998a_Li}.  H$\alpha$ EW = $-$4.00\,\AA; Li EW = 350\,m\AA, b: H$\alpha$ EW = 0.6\,\AA\ \citep{1995a_Mason}, c: $\log g = 4.39$; [Fe/H] = $-$0.01 \citep{2012a_Pinsonneault}.
\end{flushleft}
\label{T_Temp_EW_Abun_LYS}
\end{table*}}

%% file: Ages_LYS.tex
{\centering
\begin{table}
  \caption{Age estimates in Myr for the entire observed sample, based on Gyrochronology, H$\alpha$, Li EW and $R\sin i$/colour.  The final age estimate is solely from the Li EW age, other age indicators are used only as supporting evidence for the Li age.}
\begin{center}
\begin{tabular}{lrrrr}
\hline
\hline
Target & Gyro  & Li EW             & H$\alpha$        & $R\sin i$ \\
       & (Myr) & (Myr)             & (Myr)            & (Myr)     \\
\hline
A      &   $< $300 &           30--150 &           $> $21 &           \\ 
B      &   $< $100 &           30--200 &           $> $17 &           \\ 
C      &   $< $100 &           30--200 &           $> $20 &           \\ 
D      &   $< $300 &           30--200 &           $> $21 &           \\ 
E      &   $< $300 &           30--200 &           $> $27 &           \\ 
F      &   $< $300 &           30--200 &           $> $24 &           \\ 
G      &   $< $500 &           30--150 &           $> $41 &           \\ 
H      &   $< $500 &           30--150 &           $> $32 &           \\ 
I      &   $< $500 &             5--30 &           35--50 &           \\ 
J      &   $< $300 &           30--150 &           $> $19 &           \\ 
K      &   $< $300 &           30--150 &           $> $22 &           \\ 
L      &   $< $300 &           30--150 &           $> $27 &           \\ 
M      &   $< $500 &          100--200 &           $> $32 &     $< $5 \\ 
N      &   $< $500 &           30--150 &          $< $189 &           \\ 
O      &   $< $300 &           30--150 &           29--39 &           \\ 
P      &   $< $300 &           30--150 &          66--127 &           \\ 
Q      &   $< $500 &           30--150 &           41--64 &           \\ 
R      &   $< $300 &           30--150 &           $> $22 &           \\ 
S      &   $< $700 &           30--150 &          $< $233 &    $< $80 \\ 
T      &   $< $500    &          100--200 &          $< $123 &           \\ 
U      &   $< $300 &           30--150 &           $> $37 &           \\ 
V      &   $< $300 &           30--150 &           $< $53 &           \\ 
W      &   $< $300 &           30--150 &           37--55 &    $< $80 \\ 
X      &   $< $300 &           30--150 &           $> $26 &           \\ 
Y      &   $< $300 &           30--150 &           $< $99 &           \\ 
Z      &   $< $500 &           30--150 &          $< $158 &           \\ 
\hline
\end{tabular}
\end{center}
\label{T_Ages_LYS}
\end{table}}

%% file: Kin_LYS.tex
{
\tiny
\begin{table*}
\caption{Kinematic data for the likely-young sample.}
\begin{center}
\begin{tabular}{lrrrrrrr}
  \hline
  \hline
Target      &       $\mu_{\alpha}$ &      $\mu_{\delta}$ &                      $\pi$ &                           $U$ &                           $V$ &                             $W$ &           $v\sin i$  \\       
                         &             (${\rm mas\,yr}^{-1}$) &            (${\rm mas\,yr}^{-1}$) &                      (mas) &          (${\rm km\,s}^{-1}$) &          (${\rm km\,s}^{-1}$) &            (${\rm km\,s}^{-1}$) & (${\rm km\,s}^{-1}$) \\
  \hline
A           &      11.9 $\pm$ 1.6  &  $-$14.0 $\pm$ 1.7  &                7.22$-$8.31 &    $-$7.4 $\pm$ 0.9 $\pm$ 0.2 &    $-$4.6 $\pm$ 0.9 $\pm$ 0.6 &     $-$10.2 $\pm$ 0.9 $\pm$ 0.4 &   8.0 $\pm$ 1.1  \\ 
B$^{\rm a}$ &      12.2 $\pm$ 1.5  &   $-$0.6 $\pm$ 1.5  &                5.60$-$5.62 &             $-$10.6 $\pm$ 1.5 &              $-$3.9 $\pm$ 1.5 &                $-$1.9 $\pm$ 1.6 &  35.5 $\pm$ 0.5  \\ 
C &      38.9 $\pm$ 1.4  &  $-$21.9 $\pm$ 1.4  &                8.08$-$9.04 &   $-$12.9 $\pm$ 1.9 $\pm$ 0.7 &   $-$21.1 $\pm$ 1.3 $\pm$ 1.2 &      $-$1.5 $\pm$ 1.3 $\pm$ 0.1 &  20.5 $\pm$ 1.1  \\ 
D &      11.9 $\pm$ 1.7  &   $-$4.2 $\pm$ 1.7  &                6.67$-$7.64 &   $-$12.3 $\pm$ 2.1 $\pm$ 0.3 &    $-$3.2 $\pm$ 1.4 $\pm$ 0.5 &      $-$2.4 $\pm$ 1.4 $\pm$ 0.1 &  18.0 $\pm$ 0.5  \\ 
E$^{\rm b}$ &      27.5 $\pm$ 2.1  &  $-$35.1 $\pm$ 2.1  &               9.48$-$11.75 &   $-$10.0 $\pm$ 0.7 $\pm$ 1.2 &   $-$16.2 $\pm$ 0.8 $\pm$ 1.6 &      $-$6.9 $\pm$ 1.0 $\pm$ 0.7 &   8.6 $\pm$ 0.5  \\ 
F &      12.4 $\pm$ 1.6  &  $-$14.6 $\pm$ 1.7  &               8.82$-$10.65 &   $-$10.2 $\pm$ 1.6 $\pm$ 0.2 &    $-$5.8 $\pm$ 1.0 $\pm$ 0.8 &      $-$5.5 $\pm$ 1.0 $\pm$ 0.2 &  14.6 $\pm$ 0.6  \\ 
G$^{\rm c}$ &      22.5 $\pm$ 2.1  &  $-$46.2 $\pm$ 2.1  &               8.78$-$11.37 &    $-$6.6 $\pm$ 0.6 $\pm$ 0.2 &   $-$21.9 $\pm$ 1.0 $\pm$ 3.0 &     $-$10.9 $\pm$ 0.9 $\pm$ 1.1 &   6.2 $\pm$ 0.5  \\ 
H &      27.2 $\pm$ 1.3  &  $-$52.5 $\pm$ 1.3  &               9.76$-$12.34 &   $-$13.7 $\pm$ 0.6 $\pm$ 0.3 &   $-$23.8 $\pm$ 0.6 $\pm$ 2.9 &      $-$5.6 $\pm$ 0.6 $\pm$ 0.5 &   5.4 $\pm$ 0.2  \\ 
I$^{\rm d}$ &      10.0 $\pm$ 1.2  &  $-$48.0 $\pm$ 1.2  &               6.46$-$12.74 &   $-$11.5 $\pm$ 0.5 $\pm$ 0.6 &   $-$25.4 $\pm$ 0.7 $\pm$ 8.2 &     $-$11.9 $\pm$ 0.7 $\pm$ 3.3 &   6.5 $\pm$ 0.4  \\ 
J &   $-$17.1 $\pm$ 2.2  &  $-$50.6 $\pm$ 2.1  &                7.87$-$8.55 &   $-$12.5 $\pm$ 0.9 $\pm$ 0.2 &   $-$29.5 $\pm$ 1.2 $\pm$ 1.2 &      $-$5.3 $\pm$ 1.1 $\pm$ 0.5 &   5.8 $\pm$ 1.2  \\ 
K &   $-$16.9 $\pm$ 1.6  &      7.4 $\pm$ 1.6  &                7.06$-$8.39 &   $-$10.7 $\pm$ 1.0 $\pm$ 0.9 &    $-$5.2 $\pm$ 1.0 $\pm$ 0.3 &                $-$5.5 $\pm$ 0.7 &   5.4 $\pm$ 0.5  \\ 
L &  $-$115.3 $\pm$ 1.4  &  $-$69.2 $\pm$ 1.5  &              18.20$-$22.50 &   $-$15.1 $\pm$ 0.4 $\pm$ 1.4 &   $-$27.6 $\pm$ 0.4 $\pm$ 3.0 &       $-$5.5 $\pm$0.5 $\pm$ 0.2 &   4.3 $\pm$ 1.2  \\ 
M &   $-$35.7 $\pm$ 1.9  &  $-$17.5 $\pm$ 1.9  &              11.02$-$11.08 &              $-$8.9 $\pm$ 0.8 &             $-$21.4 $\pm$ 0.8 &               $-$11.0 $\pm$ 0.6 &  20.4 $\pm$ 1.3  \\ 
N &      39.4 $\pm$ 1.6  &  $-$37.7 $\pm$ 1.6  &              25.25$-$33.18 &   $-$25.0 $\pm$ 0.4 $\pm$ 1.0 &    $-$37.2 $\pm$0.4 $\pm$ 0.1 &      $-$52.6 $\pm$0.5 $\pm$ 0.7 &   8.1 $\pm$ 0.9  \\ 
O &      16.6 $\pm$ 1.7  &   $-$7.5 $\pm$ 1.7  &                7.45$-$9.42 &    $-$8.3 $\pm$ 1.0 $\pm$ 0.5 &    $-$23.7 $\pm$1.5 $\pm$ 0.3 &      $-$1.3 $\pm$ 1.1 $\pm$ 1.0 &  14.6 $\pm$ 0.5  \\ 
P$^{\rm e}$ &  $-$135.0 $\pm$ 1.1  &     55.2 $\pm$ 1.1  & 21.71 $\pm$ 1.64 &   $-$20.2 $\pm$ 0.2 $\pm$ 0.9 &    $-$5.4 $\pm$ 0.2 $\pm$ 0.2 &      $-$1.4 $\pm$ 0.7 $\pm$ 0.2 &  19.6 $\pm$ 0.4  \\ 
Q        &   $-$62.8 $\pm$ 1.2  &  $-$72.5 $\pm$ 1.2  &              17.57$-$22.84 &    $-$6.9 $\pm$ 0.3 $\pm$ 0.2 &   $-$23.7 $\pm$ 0.3 $\pm$ 3.0 &       $-$5.5 $\pm$0.3 $\pm$ 0.1 &  16.2 $\pm$ 0.5  \\ 
R &      28.2 $\pm$ 1.4  &  $-$12.4 $\pm$ 1.4  &                8.25$-$9.76 &    $-$3.0 $\pm$ 0.7 $\pm$ 0.8 &    $-$8.1 $\pm$ 0.7 $\pm$ 0.5 &      $-$28.8 $\pm$0.8 $\pm$ 1.0 &  22.3 $\pm$ 0.4  \\ 
S &    $-$4.0 $\pm$ 1.7  &     76.3 $\pm$ 1.6  &              21.69$-$28.66 &   $-$32.1 $\pm$ 0.3 $\pm$ 1.7 &   $-$14.7 $\pm$ 0.3 $\pm$ 1.0 &     $-$13.4 $\pm$ 0.3 $\pm$ 0.6 &  16.1 $\pm$ 0.5  \\ 
T &   $-$37.3 $\pm$ 2.5  &     63.4 $\pm$ 2.4  &              24.10$-$24.19 &             $-$22.2 $\pm$ 0.5 &             $-$18.0 $\pm$ 0.5 &                $-$5.7 $\pm$ 0.5 &  16.8 $\pm$ 0.5  \\ 
U &       5.0 $\pm$ 1.5  &     63.2 $\pm$ 1.7  &              14.71$-$18.86 &   $-$25.0 $\pm$ 0.5 $\pm$ 1.9 &    $-$3.1 $\pm$ 0.5 $\pm$ 1.2 &       $-$5.4 $\pm$0.5 $\pm$ 0.4 &  18.3 $\pm$ 0.5  \\ 
V &   $-$22.4 $\pm$ 3.1  &      4.1 $\pm$ 3.1  &                7.78$-$9.95 &   $-$10.8 $\pm$ 1.6 $\pm$ 0.1 &   $-$0.6 $\pm$ 1.1 $\pm$ 27.0 &       $-$0.2 $\pm$1.6 $\pm$ 1.4 &  18.8 $\pm$ 0.5  \\ 
W &      30.9 $\pm$ 1.5  &      7.1 $\pm$ 1.5  &              16.61$-$21.26 &   $-$12.1 $\pm$ 0.4 $\pm$ 0.6 &   $-$29.2 $\pm$ 0.5 $\pm$ 0.3 &     $-$13.8 $\pm$ 0.4 $\pm$ 0.8 &  37.0 $\pm$ 0.4  \\ 
X &      18.9 $\pm$ 1.5  &   $-$7.0 $\pm$ 1.5  &                7.31$-$8.98 &    $-$7.5 $\pm$ 0.7 $\pm$ 0.6 &     $-$4.5 $\pm$0.8 $\pm$ 0.3 &       $-$8.5 $\pm$0.8 $\pm$ 1.0 &  16.5 $\pm$ 0.5  \\ 
Y &      38.1 $\pm$ 1.3  &     16.2 $\pm$ 1.3  &              11.31$-$14.84 &   $-$15.1 $\pm$ 0.5 $\pm$ 2.0 &    $-$3.3 $\pm$ 0.5 $\pm$ 0.2 &      $-$0.4 $\pm$ 0.5 $\pm$ 0.2 &  21.9 $\pm$ 0.4  \\ 
Z &     105.7 $\pm$ 2.0  &  $-$85.7 $\pm$ 1.4  &              17.58$-$23.06 &   $-$11.4 $\pm$ 0.5 $\pm$ 1.5 &    $-$25.6 $\pm$0.3 $\pm$ 2.6 &     $-$18.1 $\pm$ 0.4 $\pm$ 3.2 &  15.8 $\pm$ 0.5  \\ 
\hline
\end{tabular}
\label{T_Kin_LYS}
\end{center}
\begin{flushleft}
a: $v\sin i = 34.0\,$km/s \citep{2005a_Glebocki}, b: Likely $\alpha$ Persei member \citep{2000a_Hoogerwerf}, c:  Pleiades member \citep{2014a_Sarro}, $v\sin i = 11.0 \pm 3.00\,$km/s \citep{2005a_Glebocki}, $v\sin i = 12.2 \pm 1.0\,$km/s \citep{2009a_Mermilliod}, d: $v\sin i$ = 15.6 $\pm$ 1.0\,km/s \citep{2012a_Biazzo}.  e: Parallax from \cite{2012a_de_Bruijne}.
\end{flushleft}
\end{table*}}

%% file: UVW_MG.tex
{\tiny
\begin{table*}
\begin{center}
\caption{Properties of the nearest known MGs within 100 pc.  Column 2 is the number of MG objects with a measured parallax, from \protect\cite{2014a_Gagne}.  The age ranges used for each MG are from the following publications: TWA, Columba, Carina, Argus and ABDMG (\protect\citealt{2014a_Gagne}); Octans-Near (\protect\citealt{2013a_Zuckerman}); $\eta$~Cha (\protect\citealt{2004a_Luhman}) and $\epsilon$~Cha (\protect\citealt{2013a_Murphy}). The age ranges for BPMG (\citealt{2014a_Binks, 2014b_Malo}) and Tucana-Horologium (Tuc-Hor, \citealt{2014a_Kraus}) are solely from the `lithium depletion boundary' technique.  $UVWXYZ$ are from \protect\cite{2014a_Gagne}, except for Octans (calculated from the available data in \protect\citealt{2013a_Zuckerman}) and $\eta$ and $\epsilon$~Cha (\protect\citealt{2013a_Murphy}).  Columns 7 and 8 correspond to the convergent points in right ascension and declination, respectively.}
\begin{tabular}{lrrrrrrr}
\hline
\hline
Name           & $N$ &      Age &  Distance &                        $UVW$ &                        $XYZ$ & $\alpha_{\rm CP}$ & $\delta_{\rm CP}$ \\
               &     &          &           &               $\sigma_{UVW}$ &               $\sigma_{XYZ}$ &                   &                   \\
               &     &    (Myr) &      (pc) &         (${\rm km\,s}^{-1}$) &                         (pc) &      ($^{\circ}$) &      ($^{\circ}$) \\
\hline
TWA            & 12  &   8$-$12 &   42$-$92 &  $-$10.53, $-$18.27, $-$5.00 &       12.17, $-$43.23, 21.90 &            180.87 &          $-$79.93 \\
               &     &          &           &             3.50, 1.17, 2.15 &             6.14, 7.30, 3.06 &                   &                   \\
BPMG           & 44  &  21$-$26 &    9$-$73 &  $-$11.16, $-$16.19, $-$9.27 &      4.35, $-$5.82, $-$13.29 &             88.00 &          $-$30.16 \\
               &     &          &           &             2.06, 1.32, 1.35 &           31.43, 15.04, 7.56 &                   &                   \\
ABDMG          & 48  & 70$-$120 &   11$-$64 &  $-$7.11, $-$27.21, $-$13.82 &      $-$2.25, 2.93, $-$15.42 &             92.89 &          $-$47.73 \\
               &     &          &           &             1.39, 1.31, 2.26 &          20.10, 18.97, 15.37 &                   &                   \\
Tuc$-$Hor      & 42  &  39$-$43 &   36$-$71 &   $-$9.93, $-$20.72, $-$0.89 &    11.80, $-$20.79, $-$35.68 &            116.28 &          $-$28.80 \\
               &     &          &           &             1.55, 1.79, 1.41 &             18.57, 9.14,5.29 &                   &                   \\
Columba        & 20  &  20$-$40 &   35$-$81 &  $-$12.24, $-$21.27, $-$5.56 & $-$28.22, $-$29.74, $-$28.07 &            103.27 &          $-$29.79 \\
               &     &          &           &             1.08, 1.22, 0.94 &          13.68, 23.70, 16.09 &                   &                   \\
Carina         &  5  &  20$-$40 &   46$-$88 &  $-$10.50, $-$22.36, $-$5.84 &    15.55, $-$58.53, $-$22.95 &            104.93 &          $-$34.25 \\
               &     &          &           &             0.99, 0.55, 0.14 &            5.66, 16.69, 2.74 &                   &                   \\
Argus          & 11  &  30$-$50 &    8$-$68 &  $-$21.78, $-$12.08, $-$4.52 &     14.60, $-$24.67, $-$6.72 &             91.93 &           $-$1.24 \\
               &     &          &           &             1.32, 1.97, 0.50 &          18.60, 19.06, 11.43 &                   &                   \\
Octans-Near    & 14  & 30$-$100 &   24$-$98 &  $-$13.04, $-$3.53, $-$11.24 &        8.07, 25.32, $-$48.97 &             60.12 &           $-$4.74 \\
               &     &          &           &             1.86, 2.17, 2.00 &           3.76, 11.81, 22.83 &                   &                   \\
$\eta$~Cha     &  4  &   5$-$10 & $\sim 97$ & $-$10.20, $-$20.70, $-$11.20 &    33.40, $-$81.00, $-$34.90 &             89.86 &          $-$37.55 \\
               &     &          &           &             0.20, 0.10, 0.10 &             0.40, 1.00, 0.40 &                   &                   \\
$\epsilon$~Cha & 35  &    3$-$5 & 100$-$120 &  $-$10.90, $-$20.40, $-$9.90 &    54.00, $-$92.00, $-$26.00 &             92.48 &          $-$35.13 \\
               &     &          &           &             0.80, 1.30, 1.40 &             3.00, 6.00, 7.00 &                   &                   \\
\hline
\end{tabular}
\label{T_UVW_MG}
\end{center}
\end{table*}}

%% file: Chisq.tex
{
\begin{table*}
\centering
\caption{$\bar\chi_{\rm T}^{2}$ values for each candidate/MG match in the likely-young sample.}
 \begin{tabular}{lrrrrrrrrrrr}
\hline
\hline
Target &   TWA &       BPMG &      ABDMG &    Tuc-Hor &    Columba &     Carina &      Argus & Octans-Near & $\eta$~Cha & $\epsilon$~Cha \\
\hline
A      &  45.9  &      22.2  &      81.5  &      38.3  &      62.3  &     186.8  &      68.2  &        3.2  &     240.5  &      45.7  \\
B      &  63.1  &      38.8  &     116.9  &      29.6  &      73.4  &     638.7  &      38.8  &        7.9  &   11511.3  &      64.6  \\
C      &  16.1  &      13.5  &      18.2  &       1.1  &       6.1  &     230.8  &      27.9  &       24.3  &    1808.2  &      13.0  \\
D      &  62.1  &      37.3  &     111.8  &      31.0  &      67.2  &     346.4  &      28.5  &        6.6  &    1490.7  &      61.9  \\
E      &  10.8  &       0.9  &      13.0  &       5.9  &       3.2  &       5.1  &      17.4  &        9.4  &      13.4  &       2.7  \\
F      &  36.8  &      17.3  &      69.4  &      22.6  &      37.6  &      92.4  &      28.7  &        3.7  &     346.5  &      33.1  \\
G      &  10.4  &       3.0  &       1.4  &      12.2  &      13.6  &      12.4  &      55.8  &       12.3  &      65.6  &       9.6  \\
H      &  16.9  &       4.4  &      11.6  &       5.5  &       0.8  &       3.2  &      16.5  &       12.8  &      67.7  &       6.5  \\
I      &  14.1  &       0.6  &       3.0  &       3.5  &       1.3  &       1.4  &      19.1  &        2.4  &       1.6  &       0.3  \\
J      &  29.9  &      21.8  &      10.1  &       9.5  &       8.0  &      12.1  &      35.8  &       39.9  &      91.3  &      13.7  \\
K      &  50.9  &      24.8  &      96.3  &      28.1  &      55.5  &     264.8  &      20.9  &        3.4  &    1901.8  &      47.0  \\
L      &  18.3  &       7.3  &      10.1  &       6.9  &       2.2  &       3.9  &      11.3  &       16.8  &     176.9  &       7.1  \\
M      &  13.9  &       3.4  &       2.4  &      10.8  &       6.4  &       8.4  &      40.8  &       16.4  &       1.7  &       1.3  \\
N      & 265.1  &     361.1  &     146.2  &     403.5  &     602.5  &    1599.4  &    1039.7  &      216.4  &    5716.4  &     342.3  \\
O      &  16.8  &      18.9  &      11.6  &       1.2  &       8.9  &      10.5  &      45.6  &       37.3  &      69.7  &      13.8  \\
P      &  44.8  &      29.2  &      83.5  &      24.1  &      47.5  &     132.2  &      10.4  &       10.8  &     245.2  &      47.4  \\
Q      &   9.2  &       5.9  &       4.9  &       5.0  &       8.2  &       5.6  &      46.5  &       16.6  &     436.3  &      11.7  \\
R      &  59.8  &      60.5  &      77.1  &     107.4  &     144.6  &     310.8  &     206.9  &       30.0  &     347.7  &      82.5  \\
S      &  47.0  &      23.9  &      64.0  &      56.8  &      55.7  &     103.5  &      49.4  &       27.2  &      73.6  &      50.4  \\
T      &  25.5  &      10.1  &      39.1  &      13.5  &      12.9  &      35.8  &       3.2  &       23.2  &     161.6  &      18.3  \\
U      &  54.9  &      29.2  &      87.1  &      38.6  &      50.3  &      90.2  &       6.4  &        9.6  &     153.7  &      51.9  \\
V      &  28.0  &      25.7  &      11.2  &       3.8  &       9.9  &      16.0  &      43.0  &       43.4  &      55.8  &      15.2  \\
W      &  45.7  &      33.5  &       4.5  &      29.4  &      28.6  &      74.6  &      74.1  &       46.2  &     227.6  &      16.9  \\
X      &  52.4  &      25.8  &      96.3  &      33.7  &      65.7  &     271.0  &      41.8  &        3.3  &     832.5  &      51.4  \\
Y      &  68.8  &      46.0  &     124.9  &      32.9  &      81.3  &     495.3  &      27.3  &        9.8  &    3401.3  &      73.2  \\
Z      &  17.1  &       5.8  &       1.9  &       9.2  &       5.7  &       5.6  &      20.6  &       15.6  &       3.0  &       3.0  \\
\hline
\end{tabular}
\label{T_Chisq}
\end{table*}}

%% file: RV_Proj.tex
{
\centering
\begin{table*}
\caption{$\Delta RV$ values for all MG/candidate comparisons}
\begin{tabular}{lrrrrrrrrrr}
\hline
\hline
Target &   TWA &       BPMG & ABDMG & Tuc-Hor & Columba & Carina & Argus & Octans-Near & $\eta$~Cha & $\epsilon$~Cha \\
\hline
A      &  21.0 &        5.5 &  11.5 &    13.8 &    10.1 &   11.5 &   0.1 &         4.3 &        7.5 &            7.7 \\
B      &  17.2 &        1.4 &   6.4 &     9.8 &     5.8 &    7.2 &   3.6 &         7.6 &        3.0 &            3.3 \\
C      &  14.8 &        4.5 &   1.9 &     2.4 &     1.2 &    0.5 &  12.4 &        12.8 &        2.4 &            2.5 \\
D      &  23.6 &        3.2 &   9.0 &     9.5 &     5.9 &    7.6 &   5.1 &         4.1 &        5.0 &            4.8 \\
E      &  17.4 &        1.0 &   8.9 &     2.9 &     1.3 &    3.4 &  11.8 &        10.3 &        2.5 &            1.8 \\
F      &  22.1 &        1.2 &   6.6 &     7.1 &     3.5 &    5.2 &   7.4 &         5.4 &        2.8 &            2.6 \\
G      &  16.8 &        4.5 &   0.7 &     1.1 &     2.9 &    1.2 &  12.7 &         9.6 &        3.6 &            3.7 \\
H      &  23.2 &        1.1 &   6.0 &     3.8 &     1.1 &    2.9 &   9.3 &         2.5 &        2.4 &            1.8 \\
I      &  22.9 &        0.9 &   4.1 &     3.1 &     0.2 &    1.8 &   9.0 &         0.9 &        1.5 &            1.0 \\
J      &  19.2 &        5.2 &   9.8 &     0.4 &     1.4 &    2.8 &   5.5 &         6.5 &        6.6 &            5.2 \\
K      &   5.0 &        7.2 &  15.4 &     0.9 &     5.4 &    6.1 &   0.8 &         4.7 &       10.6 &            9.2 \\
L      &  10.9 &        4.3 &   2.7 &     4.4 &     0.4 &    0.6 &   7.2 &        10.6 &        4.1 &            3.4 \\
M      &  11.2 &        2.8 &   4.0 &     9.2 &     4.2 &    4.1 &   5.9 &         4.7 &        0.2 &            0.9 \\
N      &  59.9 &       47.9 &  40.6 &    51.8 &    47.2 &   47.2 &  48.5 &        52.7 &       44.6 &           45.3 \\
O      &   9.5 &        9.1 &   1.2 &     2.4 &     3.1 &    2.4 &   9.6 &        21.3 &        5.8 &            5.5 \\
P      &   1.6 &        6.4 &  10.5 &     1.9 &     3.1 &    3.1 &   3.5 &         8.4 &        8.3 &            7.1 \\
Q      &  14.4 &        5.1 &   5.6 &     0.6 &     4.1 &    3.0 &  10.2 &         6.8 &        5.9 &            5.6 \\
R      &  17.1 &        4.5 &   2.6 &     8.3 &     3.7 &    3.8 &   4.6 &         9.2 &        1.2 &            1.9 \\
S      &  26.8 &       13.3 &   6.4 &    15.5 &    11.4 &   11.6 &  12.2 &        19.2 &       10.1 &           10.5 \\
T      &  13.4 &        3.9 &   5.1 &     5.6 &     1.7 &    1.4 &   5.2 &        11.3 &        0.1 &            0.5 \\
U      &   8.9 &        3.4 &  11.0 &     1.8 &     5.7 &    5.7 &   4.1 &         3.3 &        7.0 &            6.5 \\
V      &  13.0 &        5.0 &   4.6 &     5.3 &     1.9 &    1.6 &   6.5 &        13.9 &        0.8 &            1.3 \\
W      &  17.5 &       13.6 &   3.0 &    11.5 &     9.4 &    8.7 &  16.1 &        24.7 &        9.1 &            9.5 \\
X      &   2.5 &        8.3 &   9.1 &    14.4 &    13.7 &   13.0 &  15.5 &         1.7 &        9.1 &           10.1 \\
Y      &   9.7 &        5.6 &  12.1 &    14.1 &    11.9 &   12.7 &   5.0 &         5.9 &        8.2 &            8.7 \\
Z      &   2.7 &        1.7 &   4.0 &     7.1 &     4.5 &    5.2 &   2.0 &        12.8 &        0.6 &            1.1 \\
\hline
\end{tabular}
\label{T_RV_Proj}
\end{table*}}

%% file: MG_Test.tex
{
\centering
\begin{table*}
\caption{The full list of candidate/MG criteria matches for the likely-young sample.  A \tm\ marks a successful match and a \xm\ denotes a failed criteria.  The columns labelled `1, 2 and 3' are the criteria listed in $\S$\ref{S_MG_Criteria}.  An object was only considered a potential member of an MG if all three were satisfied.}
\begin{tabular}{p{0.7cm}p{0.07cm}p{0.07cm}p{0.25cm}p{0.07cm}p{0.07cm}p{0.25cm}p{0.07cm}p{0.07cm}p{0.25cm}p{0.07cm}p{0.07cm}p{0.25cm}p{0.07cm}p{0.07cm}p{0.25cm}p{0.07cm}p{0.07cm}p{0.25cm}p{0.07cm}p{0.07cm}p{0.25cm}p{0.07cm}p{0.07cm}p{0.25cm}p{0.07cm}p{0.07cm}p{0.25cm}p{0.07cm}p{0.07cm}p{0.25cm}}
\hline
\hline
              & \multicolumn{3}{l}{TWA} & \multicolumn{3}{l}{BPMG} & \multicolumn{3}{l}{ABDMG} & \multicolumn{3}{l}{Tuc-Hor} & \multicolumn{3}{l}{Columba} & \multicolumn{3}{l}{Carina} & \multicolumn{3}{l}{Argus} & \multicolumn{3}{l}{Octans-Near} & \multicolumn{3}{l}{$\eta$~Cha} & \multicolumn{3}{l}{$\epsilon$~Cha} \\
Target &           1 &   2 &   3 &                  1 &   2 &   3 &             1 &   2 &   3 &               1 &   2 &   3 &               1 &   2 &   3 &              1 &   2 &   3 &             1 &   2 &   3 &                   1 &   2 &   3 &                  1 &   2 &   3 &                          1 & 2 & 3 \\
\hline
A      &         \xm & \xm & \xm &                \xm & \xm & \xm &           \xm & \xm & \tm &             \xm & \xm & \tm &             \xm & \xm & \tm &            \xm & \xm & \tm &           \xm & \tm & \tm &                 \tm & \tm & \tm &                \xm & \xm & \xm &                    \xm & \xm & \xm \\  
B      &         \xm & \xm & \xm &                \xm & \tm & \xm &           \xm & \xm & \tm &             \xm & \xm & \tm &             \xm & \xm & \tm &            \xm & \xm & \tm &           \xm & \tm & \tm &                 \xm & \xm & \tm &                \xm & \tm & \xm &                    \xm & \tm & \xm \\
C      &         \xm & \xm & \xm &                \xm & \tm & \xm &           \xm & \tm & \tm &             \tm & \tm & \tm &             \xm & \tm & \tm &            \xm & \tm & \tm &           \xm & \xm & \tm &                 \xm & \xm & \tm &                \xm & \tm & \xm &                    \xm & \tm & \xm \\
D      &         \xm & \xm & \xm &                \xm & \tm & \xm &           \xm & \xm & \tm &             \xm & \xm & \tm &             \xm & \xm & \tm &            \xm & \xm & \tm &           \xm & \xm & \tm &                 \xm & \tm & \tm &                \xm & \tm & \xm &                    \xm & \tm & \xm \\
E      &         \xm & \xm & \xm &                \tm & \tm & \xm &           \xm & \xm & \tm &             \xm & \tm & \tm &             \tm & \tm & \tm &            \xm & \tm & \tm &           \xm & \xm & \tm &                 \xm & \xm & \tm &                \xm & \tm & \xm &                    \tm & \tm & \xm \\
F      &         \xm & \xm & \xm &                \xm & \tm & \xm &           \xm & \xm & \tm &             \xm & \xm & \tm &             \xm & \tm & \tm &            \xm & \xm & \tm &           \xm & \xm & \tm &                 \tm & \xm & \tm &                \xm & \tm & \xm &                    \xm & \tm & \xm \\
G      &         \xm & \xm & \xm &                \tm & \tm & \xm &           \tm & \tm & \tm &             \xm & \tm & \tm &             \xm & \tm & \tm &            \xm & \tm & \tm &           \xm & \xm & \tm &                 \xm & \xm & \tm &                \xm & \tm & \xm &                    \xm & \tm & \xm \\
H      &         \xm & \xm & \xm &                \xm & \tm & \xm &           \xm & \xm & \tm &             \xm & \tm & \tm &             \tm & \tm & \tm &            \tm & \tm & \tm &           \xm & \xm & \tm &                 \xm & \tm & \tm &                \xm & \tm & \xm &                    \xm & \tm & \xm \\
I      &         \xm & \xm & \tm &                \tm & \tm & \tm &           \tm & \tm & \xm &             \tm & \tm & \tm &             \tm & \tm & \tm &            \tm & \tm & \tm &           \xm & \xm & \tm &                 \tm & \tm & \tm &                \tm & \tm & \tm &                    \tm & \tm & \tm \\
J      &         \xm & \xm & \xm &                \xm & \xm & \xm &           \xm & \xm & \tm &             \xm & \tm & \tm &             \xm & \tm & \tm &            \xm & \tm & \tm &           \xm & \xm & \tm &                 \xm & \xm & \tm &                \xm & \xm & \xm &                    \xm & \xm & \xm \\
K      &         \xm & \tm & \xm &                \xm & \xm & \xm &           \xm & \xm & \tm &             \xm & \tm & \tm &             \xm & \xm & \tm &            \xm & \xm & \tm &           \xm & \tm & \tm &                 \tm & \tm & \tm &                \xm & \xm & \xm &                    \xm & \xm & \xm \\
L      &         \xm & \xm & \xm &                \xm & \tm & \xm &           \xm & \tm & \tm &             \xm & \tm & \tm &             \tm & \tm & \tm &            \xm & \tm & \tm &           \xm & \xm & \tm &                 \xm & \xm & \tm &                \xm & \tm & \xm &                    \xm & \tm & \xm \\
M      &         \xm & \xm & \xm &                \tm & \tm & \xm &           \tm & \tm & \tm &             \xm & \xm & \tm &             \xm & \tm & \tm &            \xm & \tm & \tm &           \xm & \xm & \tm &                 \xm & \tm & \tm &                \tm & \tm & \xm &                    \tm & \tm & \xm \\
N      &         \xm & \xm & \xm &                \xm & \xm & \xm &           \xm & \xm & \tm &             \xm & \xm & \tm &             \xm & \xm & \tm &            \xm & \xm & \tm &           \xm & \xm & \tm &                 \xm & \xm & \tm &                \xm & \xm & \xm &                    \xm & \xm & \xm \\
O      &         \xm & \xm & \xm &                \xm & \xm & \xm &           \xm & \tm & \tm &             \tm & \tm & \tm &             \xm & \tm & \tm &            \xm & \tm & \tm &           \xm & \xm & \tm &                 \xm & \xm & \tm &                \xm & \xm & \xm &                    \xm & \xm & \xm \\
P      &         \xm & \tm & \xm &                \xm & \xm & \xm &           \xm & \xm & \tm &             \xm & \tm & \tm &             \xm & \tm & \tm &            \xm & \tm & \tm &           \xm & \tm & \tm &                 \xm & \xm & \tm &                \xm & \xm & \xm &                    \xm & \xm & \xm \\
Q      &         \xm & \xm & \xm &                \xm & \xm & \xm &           \xm & \xm & \tm &             \xm & \tm & \tm &             \xm & \tm & \tm &            \xm & \tm & \tm &           \xm & \xm & \tm &                 \xm & \xm & \tm &                \xm & \xm & \xm &                    \xm & \xm & \xm \\
R      &         \xm & \xm & \xm &                \xm & \tm & \xm &           \xm & \tm & \tm &             \xm & \xm & \tm &             \xm & \tm & \tm &            \xm & \tm & \tm &           \xm & \tm & \tm &                 \xm & \xm & \tm &                \xm & \tm & \xm &                    \xm & \tm & \xm \\
S      &         \xm & \xm & \xm &                \xm & \xm & \xm &           \xm & \xm & \tm &             \xm & \xm & \tm &             \xm & \xm & \tm &            \xm & \xm & \tm &           \xm & \xm & \tm &                 \xm & \xm & \tm &                \xm & \xm & \xm &                    \xm & \xm & \xm \\
T      &         \xm & \xm & \xm &                \xm & \tm & \xm &           \xm & \xm & \tm &             \xm & \xm & \tm &             \xm & \tm & \tm &            \xm & \tm & \tm &           \tm & \xm & \tm &                 \xm & \xm & \tm &                \xm & \tm & \xm &                    \xm & \tm & \xm \\
U      &         \xm & \xm & \xm &                \xm & \tm & \xm &           \xm & \xm & \tm &             \xm & \tm & \tm &             \xm & \xm & \tm &            \xm & \xm & \tm &           \xm & \tm & \tm &                 \xm & \tm & \tm &                \xm & \xm & \xm &                    \xm & \xm & \xm \\
V      &         \xm & \xm & \xm &                \xm & \tm & \xm &           \xm & \tm & \tm &             \tm & \xm & \tm &             \xm & \tm & \tm &            \xm & \tm & \tm &           \xm & \xm & \tm &                 \xm & \xm & \tm &                \xm & \tm & \xm &                    \xm & \tm & \xm \\
W      &         \xm & \xm & \xm &                \xm & \xm & \xm &           \xm & \tm & \tm &             \xm & \xm & \tm &             \xm & \xm & \tm &            \xm & \xm & \tm &           \xm & \xm & \tm &                 \xm & \xm & \tm &                \xm & \xm & \xm &                    \xm & \xm & \xm \\
X      &         \xm & \tm & \xm &                \xm & \xm & \xm &           \xm & \xm & \tm &             \xm & \xm & \tm &             \xm & \xm & \tm &            \xm & \xm & \tm &           \xm & \xm & \tm &                 \tm & \tm & \tm &                \xm & \xm & \xm &                    \xm & \xm & \xm \\
Y      &         \xm & \xm & \xm &                \xm & \xm & \xm &           \xm & \xm & \tm &             \xm & \xm & \tm &             \xm & \xm & \tm &            \xm & \xm & \tm &           \xm & \tm & \tm &                 \xm & \xm & \tm &                \xm & \xm & \xm &                    \xm & \xm & \xm \\
Z      &         \xm & \tm & \xm &                \xm & \tm & \xm &           \tm & \tm & \tm &             \xm & \xm & \tm &             \xm & \tm & \tm &            \xm & \xm & \tm &           \xm & \tm & \tm &                 \xm & \xm & \tm &                \tm & \tm & \xm &                    \tm & \tm & \xm \\
\hline
\end{tabular}
\label{T_MG_Test}
\end{table*}
}